\documentclass[a4paper,11pt]{article}
\usepackage{jheppub, bm, color} 
\usepackage{amssymb,amsfonts,slashed,amsthm,amsmath,graphicx, soul, empheq}
\usepackage[caption=false]{subfig}
\usepackage{float}
\usepackage{subcaption}  
\usepackage{placeins}
\begin{document}

\newcommand{\vev}[1]{ \left\langle {#1} \right\rangle }
\newcommand{\bra}[1]{ \langle {#1} | }
\newcommand{\ket}[1]{ | {#1} \rangle }
\newcommand{\eV}{ \ {\rm eV} }
\newcommand{\KeV}{ \ {\rm keV} }
\newcommand{\MeV}{\  {\rm MeV} }
\newcommand{\GeV}{\  {\rm GeV} }
\newcommand{\TeV}{\  {\rm TeV} }
\newcommand{\1}{\mbox{1}\hspace{-0.25em}\mbox{l}}
\newcommand{\Red}[1]{{\color{red} {#1}}}

\newcommand{\lmk}{\left(}  
\newcommand{\rmk}{\right)}
\newcommand{\lkk}{\left[}  
\newcommand{\rkk}{\right]}
\newcommand{\lhk}{\left \{ }  
\newcommand{\rhk}{\right \} }
\newcommand{\del}{\partial}  
\newcommand{\la}{\left\langle} 
\newcommand{\ra}{\right\rangle}
\newcommand{\half}{\frac{1}{2}}

\newcommand{\bea}{\begin{array}}
\newcommand{\eea}{\end{array}}
\newcommand{\beq}{\begin{eqnarray}}
\newcommand{\eeq}{\end{eqnarray}}
\newcommand{\eq}[1]{Eq.~(\ref{#1})}

\newcommand{\dd}{\mathrm{d}}
\newcommand{\Mpl}{M_{\rm Pl}}
\newcommand{\mg}{m_{3/2}}
\newcommand{\abs}[1]{\left\vert {#1} \right\vert}
\newcommand{\mphi}{m_{\phi}}
\newcommand{\Hz}{\ {\rm Hz}}
\newcommand{\for}{\quad \text{for }}
\newcommand{\Min}{\text{Min}}
\newcommand{\Max}{\text{Max}}
\newcommand{\Kahler}{K\"{a}hler }
\newcommand{\cphi}{\varphi}
\newcommand{\Tr}{\text{Tr}}
\newcommand{\diag}{{\rm diag}}

\newcommand{\SUf}{SU(3)_{\rm f}}
\newcommand{\Upq}{U(1)_{\rm PQ}}
\newcommand{\Zpq}{Z^{\rm PQ}_3}
\newcommand{\Cpq}{C_{\rm PQ}}
\newcommand{\ubar}{u^c}
\newcommand{\dbar}{d^c}
\newcommand{\ebar}{e^c}
\newcommand{\nubar}{\nu^c}
\newcommand{\Ndw}{N_{\rm DW}}
\newcommand{\Fpq}{F_{\rm PQ}}
\newcommand{\fpq}{v_{\rm PQ}}
\newcommand{\Br}{{\rm Br}}
\newcommand{\Lag}{\mathcal{L}}
\newcommand{\Lqcd}{\Lambda_{\rm QCD}}

\newcommand{\ji}{j_{\rm inf}} 
\newcommand{\jb}{j_{B-L}} 
\newcommand{\M}{M} 
\newcommand{\im}{{\rm Im} }
\newcommand{\re}{{\rm Re} }

\def\lrf#1#2{ \left(\frac{#1}{#2}\right)}
\def\lrfp#1#2#3{ \left(\frac{#1}{#2} \right)^{#3}}
\def\lrp#1#2{\left( #1 \right)^{#2}}
\def\REF#1{Ref.~\cite{#1}}
\def\SEC#1{Sec.~\ref{#1}}
\def\FIG#1{Fig.~\ref{#1}}
\def\EQ#1{Eq.~(\ref{#1})}
\def\EQS#1{Eqs.~(\ref{#1})}
\def\TEV#1{10^{#1}{\rm\,TeV}}
\def\GEV#1{10^{#1}{\rm\,GeV}}
\def\MEV#1{10^{#1}{\rm\,MeV}}
\def\KEV#1{10^{#1}{\rm\,keV}}
\def\blue#1{\textcolor{blue}{#1}}
\def\red#1{\textcolor{blue}{#1}}

\newcommand{\eff}{\Delta N_{\rm eff}}
\newcommand{\neff}{\Delta N_{\rm eff}}
\newcommand{\cc}{\Omega_\Lambda}
\newcommand{\Mpc}{\ {\rm Mpc}}
\newcommand{\Msolar}{M_\odot}

\def\my#1{\textcolor{blue}{#1}}
\def\MY#1{\textcolor{blue}{[{\bf MY:} #1}]}


\begin{flushright}
TU-1293\\ 
\end{flushright}

\title{
Delayed Scaling of Multi-Type Cosmic F- and D-strings in VOS Models
}

\author{Kazuto Nakamura and Masaki Yamada}
\affiliation{Department of Physics, Tohoku University, Sendai, Miyagi 980-8578, Japan}

\abstract{
We investigate the velocity-dependent one-scale (VOS) model to the case of one cosmic F-string and two D-strings as color flux tubes in pure Spin($4N$) gauge theory.
We analytically calculate the scaling string density as a function of the reconnection probabilities, and confirm our results with numerical calculations.
We also determine the timescale at which the string density reaches the scaling regime, and find that for certain values of the reconnection probability, the scaling time can become extremely large, by many orders of magnitude.
This leads to a characteristic suppression signature of the gravitational-wave signal at high frequencies, which may become observable in the frequency range of future interferometric gravitational-wave observations.
}

\emailAdd{kazuto.nakamura.q7@dc.tohoku.ac.jp}
\emailAdd{m.yamada@tohoku.ac.jp}

\maketitle
\flushbottom

\section{Introduction\label{sec:Introduction}}

Cosmic strings naturally arise in simple extensions of particle physics models, and their gravitational-wave (GW) signals offer valuable probes of physics beyond the Standard Model~\cite{Vilenkin:1981bx,Vachaspati:1984gt}.
Broadly, cosmic strings can be classified into two categories: those originating from Abelian Higgs models and those described by the effective Nambu-Goto action.
In the Abelian Higgs model, cosmic strings are formed through the spontaneous breaking of a U(1) symmetry, where their finite width and associated Nambu-Goldstone bosons play essential roles~\cite{Kibble:1976sj,Vilenkin:1982ks}.
If the U(1) symmetry is gauged, the strings become extremely thin, and their dynamics may be well approximated by the Nambu-Goto action, depending on the gauge coupling.
Their evolution has been extensively studied through both Nambu-Goto string simulations~\cite{Ringeval:2005kr,Blanco-Pillado:2011egf,Blanco-Pillado:2013qja,Blanco-Pillado:2017oxo,Blanco-Pillado:2017rnf} and analytical approaches based on the velocity-dependent one-scale (VOS) model~\cite{Kibble:1984hp,Martins:1995tg,Martins:1996jp,Martins:2000cs}.

Analogous Nambu-Goto strings also appear in string theory after brane inflation~\cite{Dvali:2003zj,Copeland:2003bj}, where fundamental (F-)strings~\cite{Dvali:2003zj,Copeland:2003bj} and D-strings (one-dimensional Dirichlet branes) \cite{Jones:2002cv,Sarangi:2002yt,Dvali:2002fi,Jones:2003da,Pogosian:2003mz,Dvali:2003zj,Copeland:2003bj} can be stretched to macroscopic scales, forming so-called \emph{cosmic superstrings}.
Quantum effects can suppress their reconnection probabilities, sometimes making them extremely small~\cite{Polchinski:1988cn,Jackson:2004zg,Hanany:2005bc}.
Consequently, cosmic superstrings exhibit distinctive network dynamics and constitute an important class of theoretical and observational models~\cite{Auclair:2019wcv}.
In particular, they have recently attracted considerable attention because the pulsar timing array (PTA) signals observed in recent years can be well explained by GW emission from Nambu-Goto strings with small reconnection probabilities~\cite{Yamada:2022aax,Yamada:2022imq,Chen:2022azo,Ellis:2023tsl,Figueroa:2023zhu,Yamada:2023thl,Ellis:2023oxs,Datta:2024bqp,Avgoustidis:2025svu}.
However, realizing such cosmic superstrings within the cosmological history of string theory remains highly nontrivial.

Recently, we have pointed out that similar types of cosmic strings can arise naturally in simple gauge theories, without invoking string theory, specifically, in pure Yang-Mills theories in four-dimensional spacetime~\cite{Yamada:2022aax,Yamada:2022imq,Yamada:2023thl}.
In pure Yang-Mills theories without light fermions, color confinement leads to the formation of color flux tubes that behave as macroscopic cosmic strings (see also Ref.~\cite{Witten:1985fp}).
According to the holographic principle, these flux tubes can be interpreted as duals of fundamental strings or wrapped D-branes (see, e.g., Refs.~\cite{Witten:1998zw,Polchinski:2000uf,Klebanov:2000hb,Maldacena:2000yy,Vafa:2000wi}).
The reconnection probability of such strings can be estimated in the large-$N$ limit~\cite{tHooft:1973alw} (see Ref.~\cite{Coleman:1985rnk} for a review), implying that these strings follow Nambu-Goto dynamics and possess small reconnection probabilities depending on the color number $N$.
Moreover, depending on the gauge structure, multiple distinct types of cosmic strings can exist, which can be understood in terms of the one-form center symmetry~\cite{Yamada:2022imq}.

In this work, we analyze the dynamics of such cosmic string networks with small reconnection probabilities using an extended version of the VOS model~\cite{Avgoustidis:2005nv}.
As the simplest yet nontrivial example, we focus on the ${\rm Spin}(4N)$ ($N \ge 2$) gauge theory in a hidden sector.
Importantly, the gauge symmetry remains unbroken in our model, so the gauge interaction becomes strong at the confinement scale.
We consider the scenario in which confinement occurs after inflation, producing macroscopic color flux tubes, i.e., cosmic strings.
Due to the structure of the one-form center symmetry, the system contains one type of F-string and two types of D-strings, forming a network of three distinct string species.%
\footnote{
Our strings differ from the non-Abelian strings discussed in Refs.~\cite{Spergel:1996ai,McGraw:1997nx,Avgoustidis:2007aa}, in which many different string types cannot intercommute or pass through one another.
Although our setup involves a non-Abelian gauge theory, we focus on the low-energy theory after confinement rather than on spontaneous gauge symmetry breaking.
}
We assume that the confinement scale $\Lambda$ is of order $10^{12}\,\mathrm{GeV}$ or lower, so that the corresponding string tension $\mu$ naturally lies at that energy scale.
While the reconnection probabilities are expected to depend on $N$, their explicit values remain theoretically uncertain.
In this paper, we therefore treat them as free parameters and explore their phenomenological implications.
Although multi-tension VOS models have been extensively developed for cosmic superstrings in string theory~\cite{Copeland:2006eh,Copeland:2006if,Copeland:2007nv,Avgoustidis:2007aa,Rajantie:2007hp,Avgoustidis:2009ke,Pourtsidou:2010gu,Sousa:2016ggw,Matsui:2020hzi,Marfatia:2023fvh,Revello:2024gwa,Avgoustidis:2025svu}, where infinitely many string types can exist, our system differs in that it contains exactly three species.

Using the VOS model with three string types and small reconnection probabilities, we investigate the evolution of the cosmic string energy density and determine the timescale for the network to reach the scaling regime.
Our results show that the time required for scaling can be significantly prolonged for certain values of the reconnection probability.
Based on these findings, we compute the loop number density and the resulting GW spectrum, showing that delayed scaling leads to a suppression of the GW amplitude in the high-frequency region.

The remainder of this paper is organized as follows. 
In Sec.~\ref{sec:theory}, we briefly explain how cosmic strings arise from pure $\mathrm{Spin}(4N)$ Yang-Mills theory and summarize their properties.
Some parts of this section are rather technical, and readers interested primarily in the phenomenology of multi-type cosmic string networks may skip it and proceed directly to the subsequent sections.
In Sec.~\ref{sec:VOS model for multiple strings}, we introduce the VOS model for a network containing three types of cosmic strings with small reconnection probabilities. 
In Sec.~\ref{sec:RD}, we analytically derive the dependence of the scaling solution on the reconnection probability, with detailed calculations provided in Appendix~\ref{sec:Analytic behavior of scaling solution in Extended VOS}.
We then numerically solve the VOS equations to confirm the analytical results and determine the timescale for the onset of scaling.
In Sec.~\ref{sec:Gravitational Waves}, we compute the GW spectrum using the loop number density obtained from the VOS equations.
Finally, our conclusions and discussions are presented in Sec.~\ref{sec:Summary and discussion}.
Throughout the main text, we employ the extended VOS model to treat the correlation length and inter-string distance independently, while the conventional VOS model with a phenomenological dependence on reconnection probability is discussed in Appendix~\ref{sec:Appendix}.

\section{Color flux tubes in Spin(4N) Yang-Mills theory
\label{sec:theory}}

We first briefly explain how cosmic strings arise from pure $\mathrm{Spin}(4N)$ Yang-Mills theory with $N = 2, 3, 4, \dots$, following Refs.~\cite{Yamada:2022aax,Yamada:2022imq}. 
The gauge group $\mathrm{Spin}(4N)$ is the simply connected double cover of $\mathrm{SO}(4N)$, satisfying
$\mathrm{SO}(4N) = \mathrm{Spin}(4N) / {\mathbb Z}_2$.
Specifically, we consider the following Lagrangian in the hidden sector:
\begin{equation}
S=\int d^4x \sqrt{-g} \frac{1}{2}\Tr [F_{\mu\nu}F^{\mu\nu}] ,
\end{equation}
where $F_{\mu\nu}$ denotes the field strength of the $\mathrm{Spin}(4N)$ gauge field.
We study a pure Yang-Mills theory without introducing any Higgs fields or fermions.
Because the gauge coupling is asymptotically free, it becomes strong at an energy scale $\Lambda$, below which confinement takes place.

We assume a cosmological history in which confinement occurs during the radiation-dominated era after inflation.
This requires the reheating temperature to be higher than the confinement scale.
As the Universe cools, the confinement phase transition proceeds at a temperature of order $\Lambda$.
For the values of $N$ relevant to our study, $N \ge 2$, this phase transition is expected to be first order and to proceed via bubble nucleation.
At the transition, regions separated by more than the typical bubble size lose causal contact.
This characteristic bubble size at completion is typically $10^{-(1\,\text{-}\,2)}$ times the Hubble length~\cite{Yamada:2023thl}.
Beyond causal horizons, macroscopic color flux tubes are formed, which at later times behave as cosmic strings.
Analogous to the Kibble-Zurek mechanism, the typical number of strings per correlation volume is $\mathcal{O}(1)$ at formation.
Thus, the initial energy density of cosmic strings is expected to be
$\rho \sim 10^{-(2\,\text{-}\,4)} \mu /H^{2}(\Lambda)$, where $H(T)$ is the Hubble parameter at temperature $T$.

In a theory with dynamical matter fields in the vector representation, a flux tube carrying a vector center charge can terminate on such particles and is therefore only metastable.
In contrast, in pure Yang-Mills theory with only adjoint gluons, there are no dynamical states that carry this charge, so flux tubes are absolutely stable as long as the corresponding one-form center symmetry remains exact.
They are stable in the sense that a long flux tube cannot be cut at any point.
However, a closed string loop can still disappear by shrinking to a point and emitting glueballs.
Thus, once they form at the phase transition, these flux tubes effectively behave as cosmic strings.

The properties of color flux tubes in confining gauge theories have been extensively studied in formal quantum field theory.
Depending on the gauge group and its global structure, multiple types of flux tubes may appear.
In pure Yang-Mills theory, their topological classification at long distances is determined by the one-form center symmetry associated with the center $Z(G)$ of the gauge group $G$~\cite{tHooft:1977nqb,tHooft:1979rtg,Gaiotto:2014kfa}.
For $\mathrm{Spin}(4N)$ gauge theory, the center is ${\mathbb Z}_2\times {\mathbb Z}_2$.
Thus there are three nontrivial one-form center charges, which we can label as $(1,0)$, $(0,1)$, and $(1,1)$.
The first two are related by the outer automorphism that exchanges the two chiral spinor representations, and we denote the corresponding strings by $D_1$ and $D_2$.
The last one is associated with the vector representation and will be denoted as the $F$-string.
The tensions of these confining flux tubes are expected to be of order the confinement scale squared and scale as 
\begin{align}
  &\mu_{D_1} = \mu_{D_2} \sim N \Lambda^2 \,,\label{D-string tension}
  \\
  &\mu_F \sim \Lambda^2 \,.\label{F-string tension}
\end{align}
Thus $\mathrm{Spin}(4N)$ Yang-Mills theory contains two types of spinor-like strings, $D_1$ and $D_2$, related by an outer automorphism, and one vector-like string, $F$.%
\footnote{
\label{footnote:2}
When $N=2$, the triality symmetry of $\mathrm{Spin}(8)$ enhances the outer automorphism group to $S_3$, which permutes the vector and the two spinor representations.
As a result, the three kinds of flux tubes are exactly degenerate in tension in pure $\mathrm{Spin}(8)$ Yang-Mills theory.
}

\begin{figure} 
    \centering
    \includegraphics[width=0.9\linewidth]{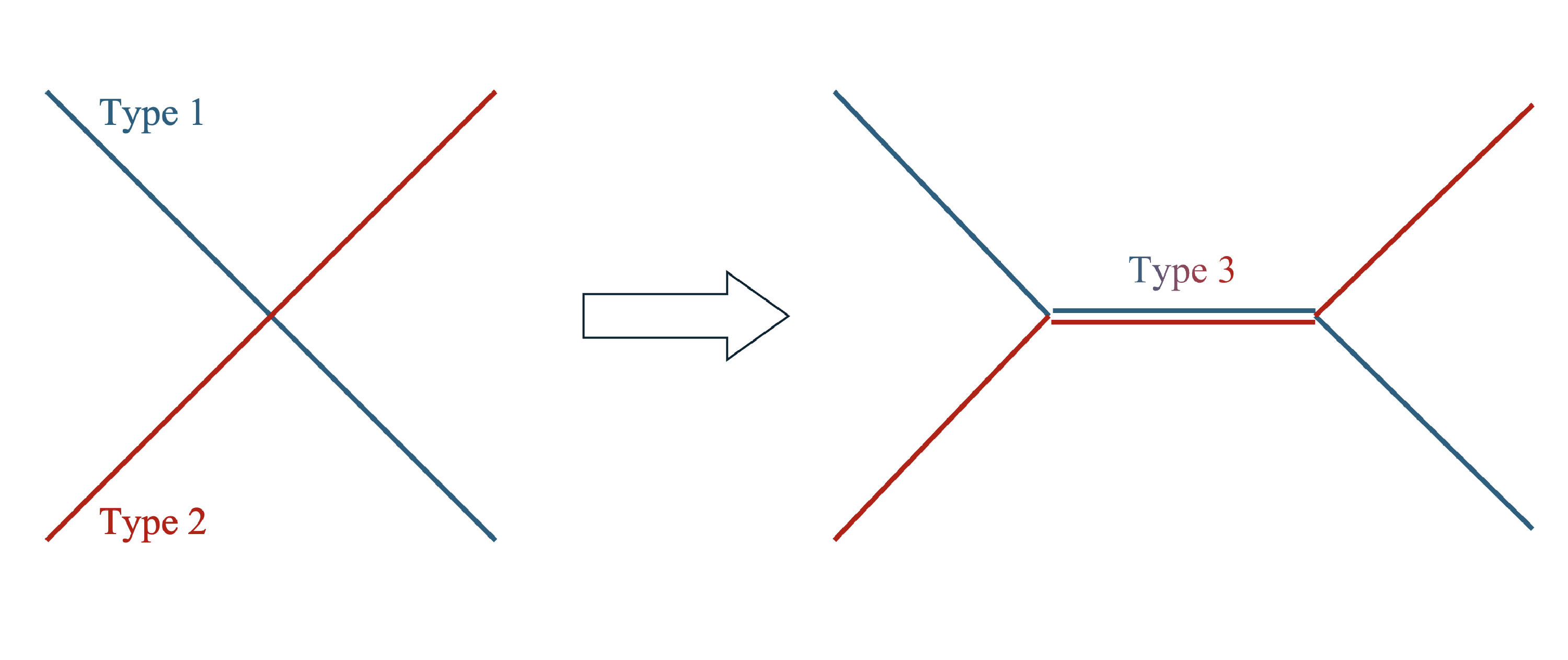}
    \caption{Schematic illustration of a zipper-type interaction.}
    \label{zipper-int}
\end{figure}

When two identical flux tubes collide, they may or may not reconnect.
At large $N$, the reconnection probability is suppressed as $\sim N^{-2}$ for $F$-strings and $\sim e^{-c N}$ for $D$-strings, as suggested by large-$N$ counting and holographic dual descriptions~\cite{Yamada:2022aax,Yamada:2022imq}.
When two different types collide, they can either pass through each other or form a third type of string via a zippering (junction-forming) process (see Fig.~\ref{zipper-int}), consistent with the conservation of the ${\mathbb Z}_2\times {\mathbb Z}_2$ charge at the junction.
The detailed dependence of these probabilities on the collision energy and angle, and the $\mathcal{O}(1)$ numerical coefficients, cannot currently be computed reliably from first-principles, and their explicit values are unknown.
The crucial point for cosmology is that these probabilities can be parametrically much smaller than unity, in close analogy to the case of cosmic superstrings in string theory~\cite{Polchinski:1988cn,Jackson:2004zg}.
For this reason, following Refs.~\cite{Yamada:2022aax,Yamada:2022imq}, we will refer to these macroscopic color flux tubes as cosmic F- and D-strings realized in pure Yang-Mills theory, emphasizing that they have superstring-like properties despite not originating from fundamental superstrings.

In the following analysis, we treat the reconnection and zippering probabilities as free parameters and investigate the statistical dynamics of the three types of cosmic strings.
Because of the exchange symmetry between $D_1$ and $D_2$, they share identical reconnection and zippering probabilities, 
as well as identical tensions.

\section{Extended VOS model for multiple string species\label{sec:VOS model for multiple strings}}

As mentioned in Sec.~\ref{sec:theory}, our interest lies in systems containing multiple types of cosmic strings with small reconnection probabilities.
In Sec.~\ref{sec:generalVOS}, we present the general form of the VOS equations for networks composed of multiple string types.
In Sec.~\ref{sec:loop}, we describe the effect of loop production following Ref.~\cite{Avgoustidis:2005nv} (see also Ref.~\cite{Yamada:2022imq}), where the correlation length and inter-string distance are treated separately.
In Sec.~\ref{sec:interaction}, we discuss the interactions between different string types.
Finally, in Sec.~\ref{sec:summaryVOS}, we summarize this section and explicitly write down the complete set of VOS equations.

\subsection{General form of the VOS equations}
\label{sec:generalVOS}

The statistical properties of cosmic string networks can be studied using the VOS model, which describes the time evolution of the energy density of long strings, $\rho$, and the root-mean-square velocity of long string segments, $v$.

In a system with multiple string species, zipper-type interactions can occur when two distinct strings collide and form a third string type between them, as illustrated in Fig.~\ref{zipper-int}.
This process has been incorporated in previous studies of cosmic superstring networks, where F-strings and D-strings can bind together through such interactions~\cite{Avgoustidis:2007aa,Avgoustidis:2009ke,Pourtsidou:2010gu,Sousa:2016ggw,Matsui:2020hzi,Marfatia:2023fvh,Avgoustidis:2025svu}.
Although our focus is not on cosmic superstrings but on macroscopic color flux tubes, their dynamics are analogous in the sesse that multiple string types coexist and tend to bind upon collision, with reconnection probabilities smaller than unity.
Hence, we adopt the VOS model developed for cosmic superstrings to describe our system.

Through a zipper-type interaction, a third string type is created by consuming portions of the colliding strings' lengths, thereby gaining the corresponding energy.
For the original two strings, this represents an energy loss, while the difference between the initial and final energies is converted into the kinetic energy of the newly formed string.

The evolution equations for the energy densities of the three string types can be formally written as~\cite{Avgoustidis:2007aa,Avgoustidis:2009ke,Pourtsidou:2010gu}:
\begin{align}
     \dot{\rho}_i &= -2H(1+v_i^2)\rho_i + \lmk \dot{\rho}_i \rmk_{\rm loop} 
     + \lmk \dot{\rho}_i \rmk_{i,j\rightarrow k}
     + \lmk \dot{\rho}_i \rmk_{k,i\rightarrow j}
     + \lmk \dot{\rho}_i \rmk_{j,k\rightarrow i} \,,
     \label{eqev d origin}
\\
    \dot{{v}}_i &= (1-{v}_i^2)\left(\frac{k(v_i)}{R_i}-2H{v}_i \right)
    + \lmk \dot{v}_i \rmk_{\rm int} \,,
    \label{eveq of vel}
\end{align}
where $H$ is the Hubble parameter and $R_i$ denotes the average radius of curvature of strings of type $i$.
The index $i$ refers to the string type ($i=1,2,3$), and $i,j,k$ are mutually distinct.
The momentum parameter $k(v_i)$ is given by an analytic fit to simulation data~\cite{Martins:2000cs}:
\begin{align}
    k(v)&=\frac{2\sqrt{2}}{\pi}\frac{1-8v^6}{1+8v^6} \,.
    \label{momentum parameter}
\end{align}

In Eq.~\eqref{eqev d origin}, the first term represents the dilution of energy density due to Hubble expansion, $(\dot{\rho})_{\rm loop}$ accounts for energy loss via loop formation, and the last three terms describe the zipper-type interactions.
For instance, $\lmk \dot{\rho}_i \rmk_{i,j\rightarrow k}$ denotes the change in $\rho_i$ due to collisions between strings of type $i$ and $j$ forming a type-$k$ string.
We have $\lmk \dot{\rho}_i \rmk_{i,j\rightarrow k}, \lmk \dot{\rho}_i \rmk_{k,i\rightarrow j} \le 0$ and $\lmk \dot{\rho}_i \rmk_{j,k\rightarrow i} \ge 0$. 
In Eq.~\eqref{eveq of vel}, the term proportional to $1/R_i$ represents acceleration due to string curvature, the term with $H$ accounts for Hubble damping, and $\lmk \dot{v}_i \rmk_{\rm int} $ describes velocity changes induced by zippering interactions.

We also define the inter-string distance $L$.
For a moving string segment of velocity $v$ occupying a volume $L^3$, the energy density of long strings can be expressed as
\begin{align}
\rho = \frac{\mu}{L^2},
\label{eq:rhotoL}
\end{align}
where $\mu$ is the string tension.
Here, one power of $L$ arises because the typical length of a long string within a volume $L^3$ is of order $L$.

The radius of curvature $R$ is expected to be of the same order as the correlation length $\xi$, and we therefore assume $R = \xi$ throughout.

\subsection{Self-interaction and loop production rate}
\label{sec:loop}

Small string loops are generated through self-intercommutation events.
In the VOS model, the energy density of long strings is separated from that of small loops, whose lengths are shorter than the Hubble length.
Accordingly, the energy density of long strings decreases due to loop production.
We denote this effect by $\lmk \dot{\rho} \rmk_{\rm loop}$.
For simplicity, we omit the string-type subscript in this subsection.

The energy loss through loop formation, $\lmk \dot{\rho} \rmk_{\rm loop}$, can be estimated as follows:
\begin{align}
    \lmk \dot{\rho} \rmk_{\rm loop} &= -(\text{effective reconnection probability})\notag\\
    &~~~\times(\text{energy of one generated loop})\notag\\
    &~~~\times(\text{number of strings that collide with one straight section per unit time})\notag\\
    &~~~\times(\text{number of straight sections per unit volume}).
\end{align}
Each factor can be evaluated separately as described below.

First, we consider the energy of one generated loop.
We introduce the loop production function $f(l,t)$, which specifies the distribution of loop lengths at time $t$. 
The quantity $f(l,t) dl$ represents the number of loops with lengths in the range $(l, l+dl)$ produced per reconnection event.
Hence, the typical energy of a loop generated per event is expressed as 
\begin{align}
    (\text{energy of one generated loop}) &= 
    \mu\int^{\infty}_0 l' f(l',t) dl' \label{eq:looppf}\\
    &\equiv \tilde{c} \mu\xi \,,
    \label{Energy~of~one~generated~loop}
\end{align}
where we define the loop-chopping efficiency parameter $\tilde{c}$. 
It quantifies the rate at which long strings lose energy through the production of closed loops. 
We expect the typical loop size to be of order the correlation length and assume that $\tilde{c}$ is constant.

Next, we estimate the number of straight string segments per unit volume.
Assuming each straight segment has a typical length of order the correlation length $\xi$, their number density can be related to the energy density of long strings:
\begin{align}
    (\text{number of straight sections per unit volume}) = \frac{\rho}{\mu\xi} = \frac{\mu}{L^2}\frac{1}{\mu\xi}=\frac{1}{\xi L^2}.
    \label{number~of~straight~sections~per~unit~volume}
\end{align}

\begin{figure}
    \centering
    \includegraphics[width=0.9\linewidth]{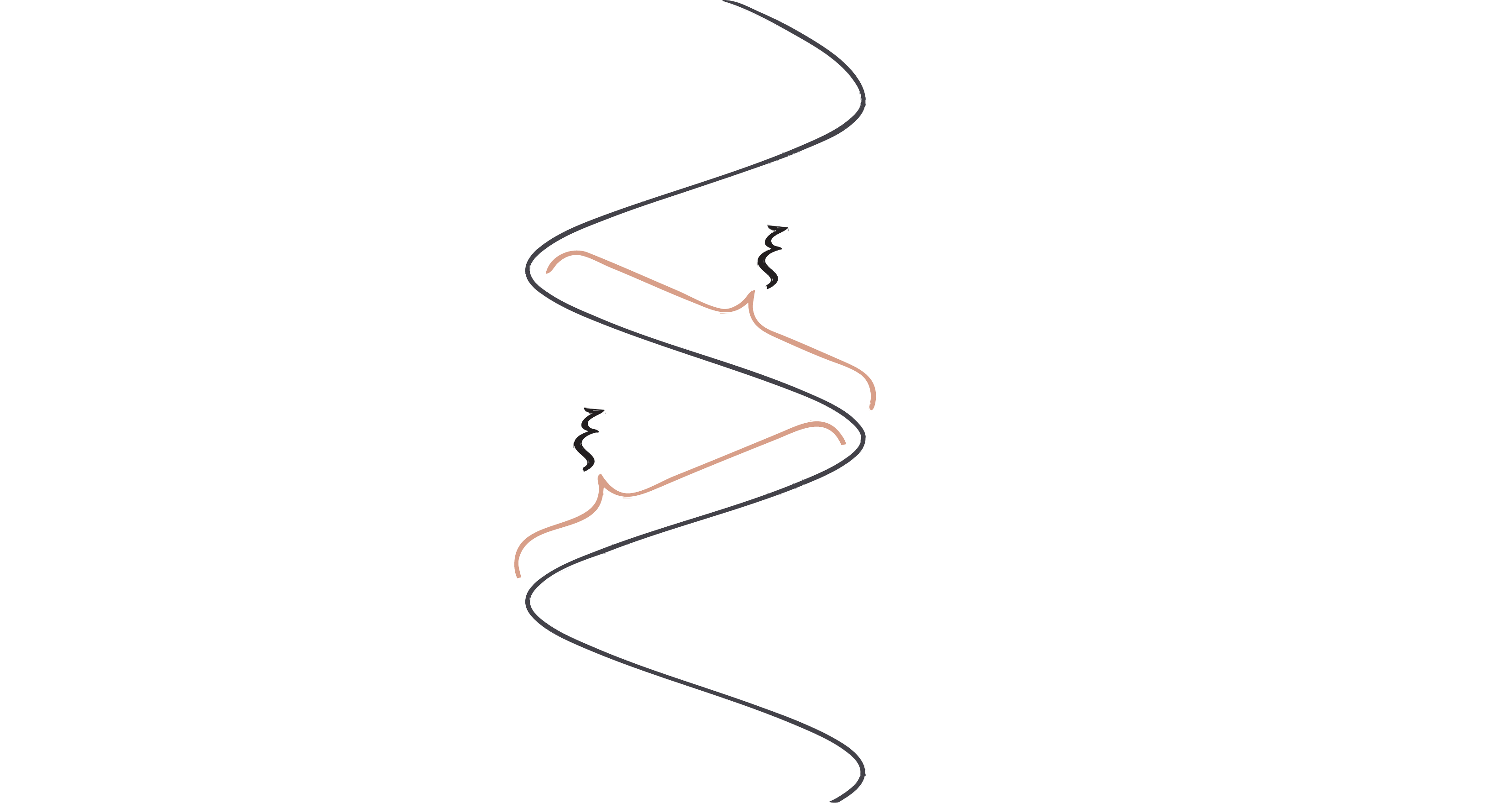}
    \caption{Schematic illustration of a straight section.}
    \label{straight section}
\end{figure}

Finally, we estimate the number of strings colliding with a straight section per unit time.
As illustrated in Fig.~\ref{straight section}, straight sections are typically separated by a distance $\sim\xi$.
After a time interval $\xi/v$, strings within a volume $\xi^3$ centered on the section, that is approximately $\xi^3 / (\xi L^2)$ strings, will collide with it.
Therefore, the number of collisions per unit time for a single straight section is
\begin{align}
    &(\text{number of strings colliding with one straight section per unit time}) =\frac{\xi^3}{\xi L^2} \frac{v}{\xi}=\frac{\xi v}{L^2}\,.
    \label{number~of~strings~that~collide~with~one~straight ~sections~in~unit~time}
\end{align}

Combining \eqref{Energy~of~one~generated~loop}, \eqref{number~of~straight~sections~per~unit~volume}, and \eqref{number~of~strings~that~collide~with~one~straight ~sections~in~unit~time}, 
the energy loss rate due to loop production can be written as
\begin{align}
     \lmk \dot{\rho} \rmk_{\text{loop}} &=-P_{\rm loop} \tilde{c} \mu\xi \frac{1}{\xi L^2} \frac{\xi v}{L^2}\notag\\
     &=-P_{\rm loop}\tilde{c}v\rho\Big(\frac{\xi}{L^2}\Big)\,,
     \label{eq:rholoop}
\end{align}
where $P_{\rm loop}$ denotes the effective reconnection probability for self-intercommutation.%
\footnote{
Numerical simulations suggest that small-scale wiggles on cosmic strings can enhance the effective reconnection probability when such structures are coarse-grained~\cite{Martins:2005es,Avgoustidis:2005nv}.
Consequently, the reconnection probability appearing in the VOS equations does not necessarily correspond directly to the theoretical value for perfectly straight strings.
We therefore adopt an effective parameterization following conventional practice.
}%
\footnote{
The reconnection probability may also depend on factors such as the collision angle and relative velocity between strings (see, e.g., Ref.~\cite{Rybak:2018mnl}).
In this work, however, we employ a simplified, constant parametrization in order to isolate and study its impact on network evolution.
}

When the reconnection probability is of order unity, it is reasonable to take $\xi = L$, in which case the extended VOS equations reduce to the standard form.
However, our interest lies in the regime of small reconnection probabilities, where it becomes necessary to treat $\xi$ and $L$ as independent quantities.
The extended VOS model presented above consistently incorporates the effects of small reconnection probabilities in agreement with numerical simulations~\cite{Martins:2005es,Avgoustidis:2005nv}.

According to Ref.~\cite{Avgoustidis:2005nv}, it is argued that self-reconnection within one Hubble time is not strongly suppressed even when the reconnection probability is small.
This is because cosmic strings are generally wiggly, possessing many left- and right-moving modes that frequently intersect within one Hubble time.
Although the reconnection probability for each individual intersection is small, the multiplicity of intersections, arising from the interactions of these left- and right-moving modes, eventually leads to reconnection events that produce string loops.
We therefore assume
\begin{align}
    \xi &= c_{\xi}t\,,
    \label{eq:xi}
\end{align}
where $c_{\xi}$ is a constant.

These considerations apply independently to all string types, so we introduce superscripts or subscripts $i$, such as $P_{\rm loop}^i$, $\mu_i$, $\tilde{c}_i$, and $c_{\xi_i}$ in place of $P_{\rm loop}$, $\mu$, $\tilde{c}$, and $c_\xi$.

\subsection{Interaction between different string types}
\label{sec:interaction}

Next, we consider interactions between different types of cosmic strings.
Through zipper-type interactions, a third string type is generated by consuming portions of the lengths of the two colliding strings and acquiring energy from those segments.

Let us specifically consider a segment of a type-1 string, of size comparable to the inter-string distance, interacting with a network of type-2 strings to produce a type-3 string.
Without loss of generality, we assume $L_1 < L_2$.
Since $L_2$ represents the typical separation between type-2 strings, there is, on average, one type-2 string within a volume $(L_2)^3$.
If $\bar{v}_{12}$ denotes the relative velocity between the type-1 and type-2 string networks, the probability that a type-1 segment collides with a type-2 string during a time interval $\delta t$ is given by
\begin{align}
    P_3 \lmk \frac{\bar{v}_{12}\delta t}{L_2} \rmk \lmk \frac{L_1}{L_2} \rmk\,,
\end{align}
where $P_3$ represents the effective reconnection probability for the production of a type-3 string through a zipper-type interaction,
and $\bar{v}_{jk}$ is the average magnitude of the relative velocity between strings of types $j$ and $k$, defined as $\bar{v}_{jk} = \sqrt{v_j^2 + v_k^2}$.
The first factor represents the probability associated with the direction of motion, while the second factor accounts for the geometrical probability of collision along the string length.
Such a collision results in the formation of a type-3 string.

For the distribution of the lengths of the produced type-3 strings, we introduce a link production function $f^3_{12}(l,t)$ and define a cross-interaction efficiency parameter $\tilde{d}^3_{12}$ by
\begin{align}
    \mu_3 \int^\infty_0 l' f^3_{12} (l',t)dl' \equiv \mu_3 \tilde{d}^3_{12} \ell^3_{12} (t) \,,
\end{align}
where $\ell^i_{jk}$ denotes the typical zipper length associated with the production of an $i$-string from the collision of $j$- and $k$-strings.
We expect the typical zipper length to be of order the correlation length $\xi$.
Using Eq.~\eqref{eq:xi}, we parameterize it as
\begin{equation}
 \ell^i_{jk} (t)  = c^i_{jk} t\,,
\end{equation}
where $c^i_{jk}$ is a constant.

On average, there are $\rho_1/(\mu_1 L_1) \times (L_2)^3 = (L_2/L_1)^3$ type-1 strings in a volume $(L_2)^3$.
Thus, the incremental energy density transferred to type-3 strings via zipper-type interactions is
\begin{align}
    (\delta {\rho_3})_{1,2\rightarrow 3} &= P_3 \tilde{d}^3_{12} \frac{\bar{v}_{12}\delta t}{L_2}\frac{L_1}{L_2}\frac{\mu_3 \ell^3_{12}(t)}{L_2^3}\frac{L_2^3}{L_1^3} \\
    &=P_k \frac{\tilde{d}^3_{12} \bar{v}_{12}\delta t\mu_3\ell^3_{12}(t)}{L_1^2L_2^2} \,.
\end{align}
Hence, the corresponding time derivative is
\begin{align}
    (\dot{\rho}_3)_{1,2\rightarrow 3} =P_3 \frac{\tilde{d}^3_{12}\bar{v}_{12}\mu_3\ell^3_{12}(t)}{L_1^2L_2^2}\,.
    \label{int link}
\end{align}
Similarly, the energy loss terms for the colliding type-1 and type-2 strings are given by
\begin{align}
    (\dot{\rho}_i)_{1,2\rightarrow 3} = - P_3 \frac{\tilde{d}^3_{12}\bar{v}_{12}\mu_i\ell^3_{12}(t)}{L_1^2L_2^2}\,,
    \label{int origin}
\end{align}
for $i = 1$ and $2$. 

Combining Eqs.~\eqref{eqev d origin}, \eqref{int link}, and \eqref{int origin},
we obtain the evolution equations for a system consisting of three string types where types 1 and 2 interact through zipper-type collisions to produce type-3 strings.

From Eq.~\eqref{int origin}, we find that if $\mu_1 + \mu_2 \neq \mu_3$, an energy imbalance arises:
\begin{align}
- \sum_i (\dot{\rho}_i)_{1,2\rightarrow 3} 
    = \tilde{d}^3_{12}\bar{v}_{12}( \mu_1 + \mu_2 - \mu_3 )\frac{\ell^3_{12}(t)}{L_1^2L_2^2}\delta t\label{remained energy}.
\end{align}
This deficit energy is converted into the kinetic energy of the newly formed type-3 string to ensure energy conservation, resulting in a change in its velocity.

Following Ref.~\cite{Avgoustidis:2007aa}, we estimate the compensation term in the velocity equation.
Since the energy density of long strings is proportional to the Lorentz factor $1/\sqrt{1-v_3^2}$
a change in the velocity induces a change in energy density according to
\begin{align}
    \frac{\partial\rho_3}{\partial v_3}\frac{dv_3}{dt}=\frac{v_3}{1-v_3^2}\rho\dot{v}_3.
    \label{acc energy}
\end{align}
This must match $(-\sum_i (\dot{\rho}_i)_{1,2\rightarrow 3})$ by energy conservation.
Accordingly, the evolution equation for the velocity of type-3 strings includes an interaction term of the form
\begin{align}
    \lmk \dot{v}_3 \rmk_{\rm int} &= B \frac{1-v_3^2}{\rho_3 v_3}\times \tilde{d}^3_{12} \bar{v}_{12} (\mu_1+\mu_2-\mu_3)\frac{\ell^3_{12}(t)}{L_1^2L_2^2}\notag\\[5pt]
    &=(1-v_3^2)\lmk B \tilde{d}^3_{12}\frac{\bar{v}_{12}}{v_3}\frac{\mu_1+\mu_2-\mu_3}{\mu_3}\frac{L_3^2\ell^3_{12}(t)}{L_1^2L_2^2}\rmk,
    \label{eq:vint}
\end{align}
where we indlude a phenomenological parameter $B$ ($0\le B \le 1$) that represents the fraction of energy that is radiated away through zipping interaction. 
For $B=1$, all excess energy is transferred into kinetic energy, while for $B=0$ all excess energy is radiated away. 
Since this value of $B$ does not change significantly the overall dynamics of the string network~\cite{Avgoustidis:2009ke}, we simply set $B=0$ throughout this paper, as we will see later.


\subsection{Explicit form of the VOS equations}
\label{sec:summaryVOS}

Combining Eqs.~\eqref{eq:rholoop}, \eqref{int link}, \eqref{int origin}, and \eqref{eq:vint},
the evolution equations for the energy density and velocity of strings of type~$i$ can be written as
\begin{align}
    \dot{\rho}_i &= -2H(1+v_i^2)\rho_i - P_{\rm loop}^i \tilde{c}_i v_i\rho_i\frac{\xi_i}{L_i^2}-P_k\frac{\tilde{d}^k_{ij}\bar{v}_{ij}\mu_i\ell^k_{ij}(t)}{L_i^2L_j^2}\notag\\
    &~~~~-P_j\frac{\tilde{d}^j_{ki}\bar{v}_{ki}\mu_i\ell^j_{ki}(t)}{L_k^2L_i^2}+P_i\frac{\tilde{d}^i_{ij}\bar{v}_{jk}\mu_i\ell^i_{jk}(t)}{L_j^2L_k^2}\,,\\
     \dot{{v}}_i &= (1-{v}_i^2)\left(\frac{k(v_i)}{\xi_i}-2H{v}_i+ B P_i\tilde{d}^i_{jk}\frac{\bar{v}_{jk}}{v_i}\frac{\mu_j+\mu_k-\mu_i}{\mu_i}\frac{L_i^2\ell^i_{jk}(t)}{L_j^2L_k^2}\right)\,,
\end{align}
where $i, j, k \in \{1,2,3\}$ are mutually distinct ($i \neq j \neq k $). 
The parameter $\tilde{d}^i_{jk}$ and $\ell^i_{jk}(t)$ denote 
the cross-intersection efficiency and the characteristic length of the link associated with the production of an $i$-string from the collision of $j$- and $k$-strings, respectively.

It is convenient to express $L_i(t)$ in the form
\begin{equation}
 L_i(t)=\gamma_i(t)t\,,
 \label{eq:gamma}
\end{equation}
where $\gamma_i(t)$ becomes constant in the scaling regime.
The corresponding evolution equation for $\gamma_i$ is then obtained as
\begin{align}
    \frac{\dot{\gamma}_i}{\gamma_i}&=\frac{1}{2t}\Bigg(2Ht(1+v^2)-2+P_{\rm loop}^i \frac{c_{\xi_i}\tilde{c}_iv_i}{\gamma_i^2}+P_k\tilde{d}^k_{ij}\bar{v}_{ij}\frac{1}{\gamma_j^2}\frac{\ell^k_{ij}(t)}{t}\notag\\
    &~~~~+P_j\tilde{d}^j_{ki}\bar{v}_{ki}\frac{1}{\gamma_k^2}\frac{\ell^j_{ki}(t)}{t}-P_i\tilde{d}^i_{jk}\bar{v}_{jk}\frac{\gamma_i^2}
    {\gamma_j^2\gamma_k^2}\frac{\ell^i_{jk}(t)}{t}\Bigg)\,,\label{eveq of gamma with eff}
\end{align}
while the velocity evolution equation becomes
\begin{equation}
    \dot{v}_i =\frac{1-v_i^2}{t}\left(\frac{k(v_i)}{\xi_i}t-2Hv_it+B P_i\tilde{d}^i_{jk}\frac{\bar{v}_{jk}}{v_i}\frac{\mu_j+\mu_k-\mu_i}{\mu_i}\frac{\gamma_i^2}{\gamma_j^2\gamma_k^2}\frac{\ell^i_{jk}(t)}{t}\right)\,,\label{eveq of v with eff}
\end{equation}
with 
\begin{align}
    \xi_i &= c_{{\xi}_i}t\,,\label{eveq of correlation length} \\
     \ell^i_{jk} (t) &= c^i_{jk} t\,,
     \label{eq:lijk}
     \\
      \bar{v}_{jk} &=\sqrt{v^2_j+v_k^2}\,,
\end{align}
where again $i, j, k$ are mutually distinct indices.

Here we emphasize that the evolution equation depends on the string tensions $\mu_i$ only through the last term within the parentheses of the velocity equation.
As we assume, it is often taken that $B = 0$, in which case the $\mu_i$ dependence disappears entirely from the combined equations, and the number density of long strings becomes independent of the string tensions.
Note, however, that the dependence on the string tensions becomes important when we compute the GW signal produced by cosmic string loops.

The above equations are motivated by both qualitative arguments and numerical results
(see also Refs.~\cite{Avgoustidis:2005nv,Yamada:2022imq}).
In the literature, an alternative formulation of the VOS model is often adopted,
where the dependence on reconnection probability is parameterized phenomenologically to fit simulation data under the assumption of $\xi_i = L_i$. Given the substantial uncertainties in numerical results, both formulations can provide consistent descriptions within current accuracy.
For completeness, we analyze the conventional form of the VOS model, frequently used in the literature,
in Appendix~\ref{sec:Appendix} for comparison.

\section{Analytic and numerical solutions in the radiation-dominated era}
\label{sec:RD}

We analyze the evolution of the cosmic string network by solving Eqs.~\eqref{eveq of gamma with eff} and~\eqref{eveq of v with eff} both analytically and numerically.
In this section, we assume a radiation-dominated (RD) universe, where the Hubble parameter is given by $H = 1/(2t)$, for simplicity.

Throughout this work, we take the $\mathcal{O}(1)$ numerical coefficients to be
\begin{align}
    \tilde{c}_i&=\tilde{c}\,,
\label{eq:tildec}
\\
    c_{\xi_i}&=c_{\xi}\,,
\label{simplification2 for e}
\\
    \tilde{d}^i_{jk} &= \tilde{d} \,,
\label{simplification1 for e}
\\
    c^i_{jk} &= c_{\xi}\,,
\label{simplification3 for e}
\end{align}
for all $i,j,k$. 
In our numerical calculations, we adopt $\tilde{c} = \tilde{d} = 0.23$, which fits simulations for both the RD and matter-dominated (MD) eras~\cite{Martins:2003vd} (see also Refs.~\cite{Bennett:1989yp,Allen:1990tv,Martins:1996jp,Martins:2000cs}).
In addition, the coefficient $c_{\xi}$ is chosen as
\begin{align}
    c_\xi  
= \begin{cases}
0.27 & \text{in RD}, \\
0.62 & \text{in MD},
\end{cases}
\label{eq:cxi}
\end{align}
so as to reproduce the standard VOS scaling solution in the case of $P_{\rm loop} = 1$.
We also assume $B = 0$ for simplicity, in which case the dependence on the string tension does not appear in the evolution equations for long strings.

We also note that in the ${\rm Spin}(4N)$ ($N \ge 2$) Yang--Mills theory considered here, two of the cosmic string species possess a symmetry.
Thus, they share the same tension and reconnection probability,
\begin{align}
    &\mu_1 = \mu_2 \\
    &P_{\rm loop}^1 = P_{\rm loop}^2 \\
     &P_1 =P_2\label{simplification4 for e}\,,
\end{align}
and evolve identically, satisfying $\gamma_1 = \gamma_2$.

For simplicity, 
we further assume 
\begin{align}
&P_{\rm loop}^i = P_{\rm loop} \label{eq:Ploopi}
\end{align}
for all $i$.
This relation is expected to hold in the Spin($8$) model due to its triality symmetry (see footnote~\ref{footnote:2}). 
Although it does not generally hold in the Spin($4N$) model with $N \ge 3$,
we adopt it here for simplicity, as it is expected to provide a good approximation when $N$ is not too large.

\subsection{Analytic behavior of the scaling solution
\label{sec:Analytic behavior of scaling solution}}

Before numerically solving the extended VOS equations,
we first derive an approximate analytic solution in the scaling regime, where $\dot{\gamma}_i = \dot{v}_i = 0$ for all $i$.
In particular, we examine how the scaling solution depends on the reconnection probabilities.

As we assume $B=0$
for simplicity, the terms involving the effective reconnection probabilities are absent in Eq.~\eqref{eveq of v with eff}.
Under this approximation, all string types share the same velocity, since they satisfy identical velocity equations.

The detailed derivation is presented in Appendix~\ref{sec:Analytic behavior of scaling solution in Extended VOS}.
Here, we summarize the asymptotic behavior of the scaling parameters $\gamma_1$ and $\gamma_3$:
\begin{align}
\gamma_1 &=\mathcal{O}(1)\times
\begin{cases}
\sqrt{P_{\text{loop}}} & \text{for } P_3 \ll P_{\text{loop}} \vspace{0.3cm}
\\
\sqrt{P_3} & \text{for }
P_{\text{loop}} \ll P_3
\end{cases}\label{gamma1 behavior for e}
\\
\notag\\
\gamma_3 &=\mathcal{O}(1)\times
\begin{cases}
\sqrt{P_{\text{loop}}} & \text{for } 
P_1 \ll {\rm Max}[P_3, \, P_{\text{loop}}] \vspace{0.3cm}
\\
\sqrt{P_1}& \text{for }
P_\text{loop} \ll P_3 \ll  P_1 \vspace{0.3cm}
\\
\sqrt{P_1 P_\text{loop} / P_3} & \text{for }
P_3 \ll P_\text{loop}\ll P_1 \\
\end{cases}\label{gamma3 behavior for e}
\end{align}
where the $\mathcal{O}(1)$ prefactors include the parameters 
$\tilde{c}$, $c_\xi$, $\tilde{d}$, and $v$. 

When there exists a hierarchy between $P_3$ and $P_{\text{loop}}$,
the qualitative behavior significantly changes depending on whether $P_1$ is larger or smaller than ${\rm Max}[P_3, \, P_{\text{loop}}]$.
In Sec.~\ref{sec:scalingtime}, we show numerically that near this threshold,
the time required for the network to reach the scaling regime can increase significantly.

It is often convenient to consider the ratio $\gamma_3/\gamma_1$: 
\begin{align}
\frac{\gamma_3}{\gamma_1} &=\mathcal{O}(1)\times
\begin{cases}
1 & \text{for } 
P_1, P_3 \ll P_{\text{loop}} \vspace{0.3cm}
\\
\sqrt{P_{\text{loop}}/ P_3} & \text{for } 
P_1, P_{\text{loop}} \ll P_3 \vspace{0.3cm}
\\
\sqrt{P_1/P_3}& \text{for }
P_3, P_\text{loop} \ll  P_1 \\
\end{cases}\label{ratio behavior for e}
\end{align}
Note that if $P_{\rm loop} \gg P_i$ ($i = 1,2,3$), the evolution of the cosmic string network is governed almost entirely by the self-reconnection process, and each string type evolves essentially independently.
Since we assume that all self-reconnection probabilities are identical, the scaling densities satisfy
$\gamma_1 = \gamma_2 \sim \gamma_3 \sim P_{\rm loop}^{1/2}$. 
On the contrary, if $P_i$ is the largest among the reconnection probabilities, then $\gamma_i$ becomes the smallest and the energy density of the type-$i$ component dominates.
This is expected, since the zippering process preferentially produces that component.

\subsection{Numerical results for scaling behavior\label{sec:Numerical calculations}}

\begin{figure}
  \centering
  \begin{minipage}{0.48\linewidth}
  \includegraphics[width=\linewidth]{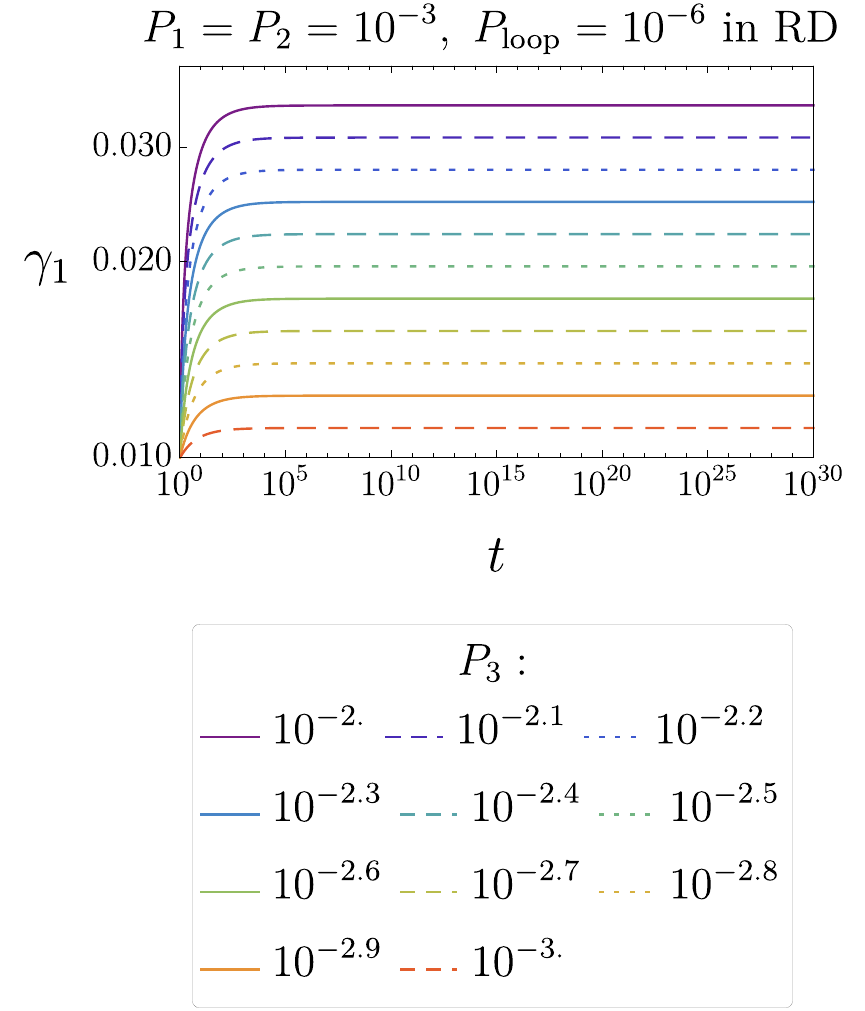}
  \end{minipage}
  \quad
  \begin{minipage}{0.48\linewidth}
  \includegraphics[width=\linewidth]{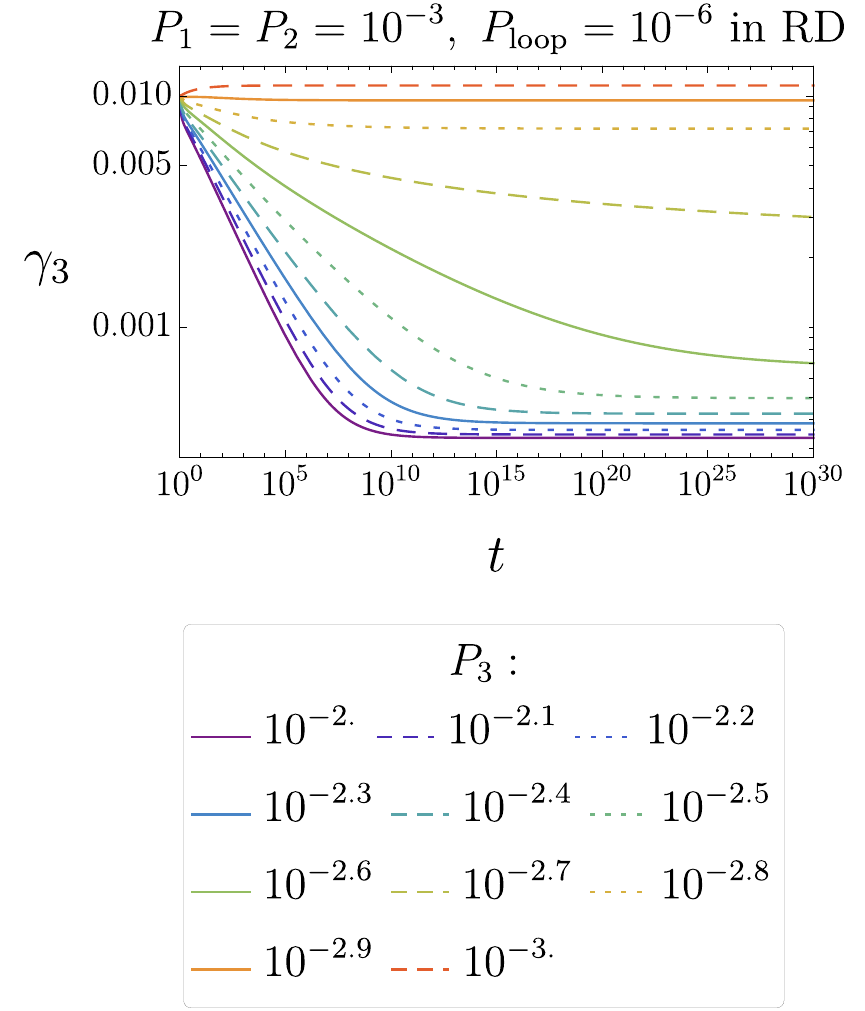}
  \end{minipage}
  \caption{
  Time evolution of $\gamma_1$ (left panel) and $\gamma_3$ (right panel) in the RD era for $P_1 = P_2 = 10^{-3}$ and $P_{\rm loop} = 10^{-6}$.
We vary $P_3$ from $10^{-2}$ to $10^{-3}$, as indicated by different colors and line styles.
  }
  \label{plot-time evolution}
\end{figure}

We numerically solve the extended VOS equations with interactions,
Eqs.~\eqref{eveq of gamma with eff}, \eqref{eveq of v with eff}, and \eqref{eveq of correlation length}.
The initial conditions for all string types are set to
\begin{align}
    \gamma_i=10^{-2},\,v_i=10^{-2}\quad\text{for }\, i=1,2,3 \,.
\end{align}
As we consider RD in this section, we have $2Ht = 1$.
Moreover, using \eqref{eveq of correlation length} and \eqref{eq:lijk}, the evolution equations for $\gamma_i$ and $v_i$ reduce to differential equations with respect to $\ln t$.
Therefore, the absolute normalization of the time variable is not meaningful, and we simply take $t = 1$ (or $\ln t = 0$) as the initial time in our numerical simulations.

Figure~\ref{plot-time evolution} shows the time evolution of $\gamma_1 = \gamma_2$ (left panel) and $\gamma_3$ (right panel) for $P_1 = P_2 = 10^{-3}$ and $P_{\rm loop} = 10^{-6}$, with different values of $P_3$.
Each $\gamma_i$ asymptotically approaches a constant value at late times, demonstrating the emergence of the scaling regime.
The parameter $\gamma_1$ reaches its scaling value within a timescale of $\mathcal{O}(10^{1\,\text{-}\,2})$,
whereas $\gamma_3$ requires a much longer time, by several orders of magnitude, for certain choices of reconnection probabilities.

\begin{figure}
  \centering
  \begin{minipage}{0.48\linewidth}
  \includegraphics[width=\linewidth]{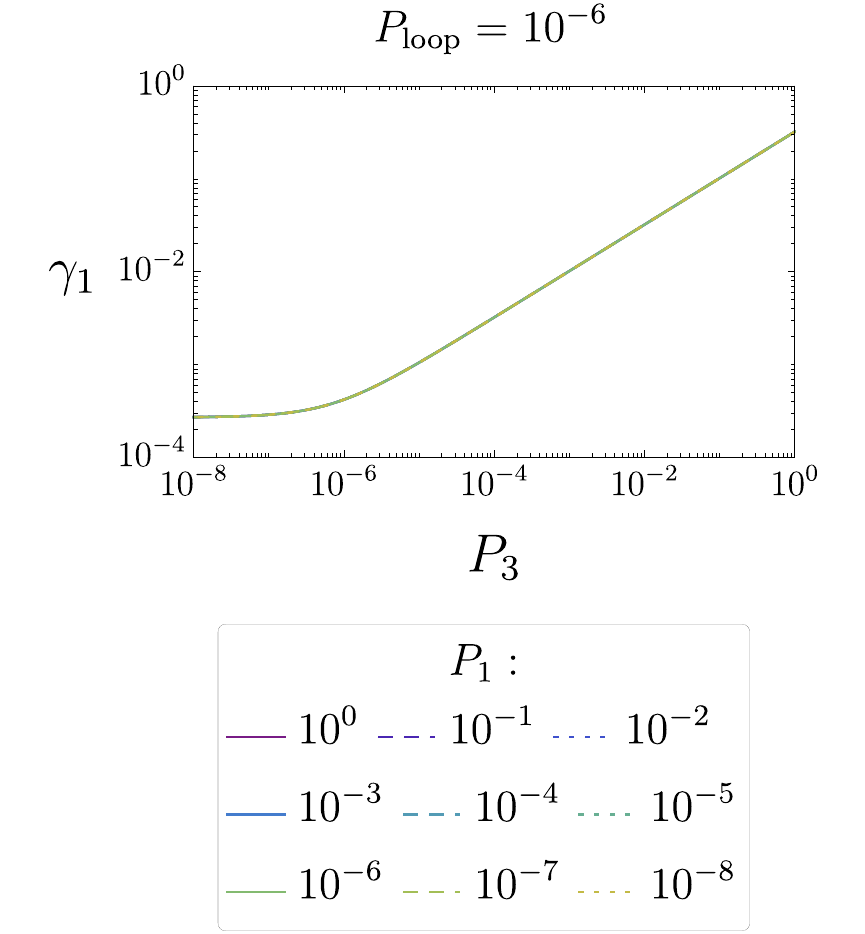}
  \end{minipage}
  \quad
  \begin{minipage}{0.48\linewidth}
  \includegraphics[width=\linewidth]{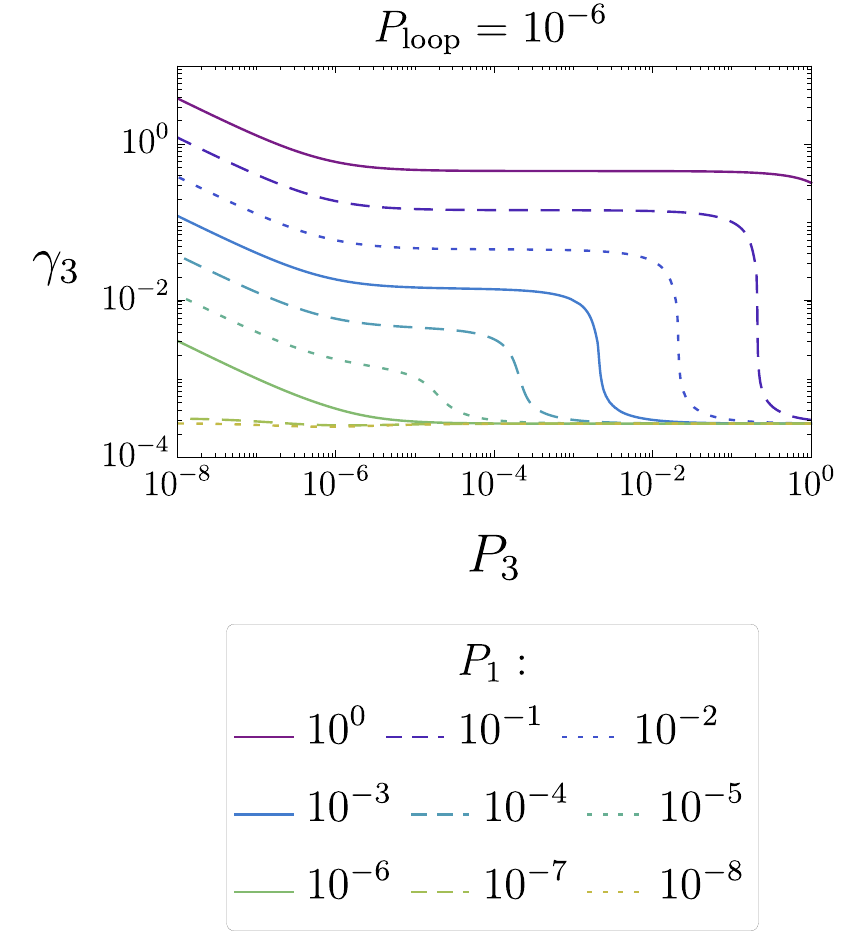}
  \end{minipage}
  \caption{
  Scaling values of $\gamma_1$ (left) and $\gamma_3$ (right) as functions of $P_3$ in the RD era,
for $P_{\rm loop} = 10^{-6}$.
We take $P_1 = P_2$ from $10^{0}$ to $10^{-8}$ (shown in different colors and line styles, from top to bottom in the right panel).
In the left panel, all curves overlap and are indistinguishable. }
  \label{plot-final value}
\end{figure}

Figure~\ref{plot-final value} shows the scaling values of $\gamma_1$ (left panel) and $\gamma_3$ (right panel) as functions of $P_3$ in the RD era, for $P_{\rm loop} = 10^{-6}$.
We take $P_1 = P_2$ to vary from $10^{0}$ to $10^{-8}$, as indicated by different colors and line styles (from top to bottom in the right panel).
In the left panel, all curves overlap and are indistinguishable.

\begin{figure}
  \centering
  \begin{minipage}{0.48\linewidth}
  \includegraphics[width=\linewidth]{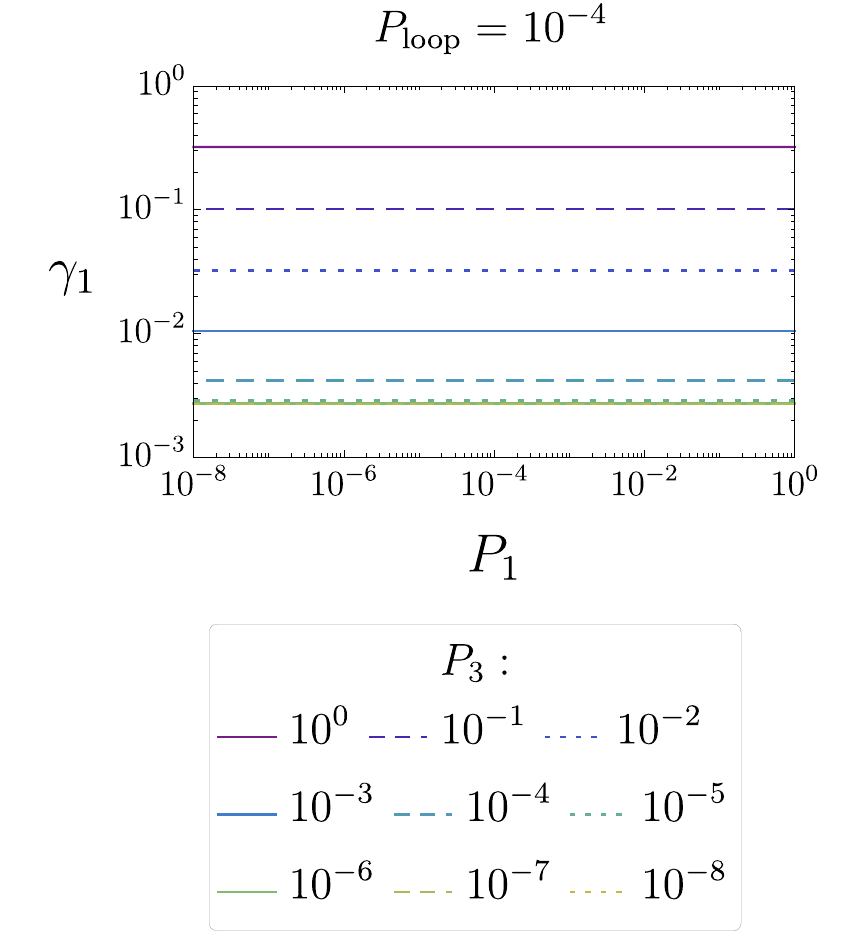}
  \end{minipage}
  \quad
  \begin{minipage}{0.48\linewidth}
  \includegraphics[width=\linewidth]{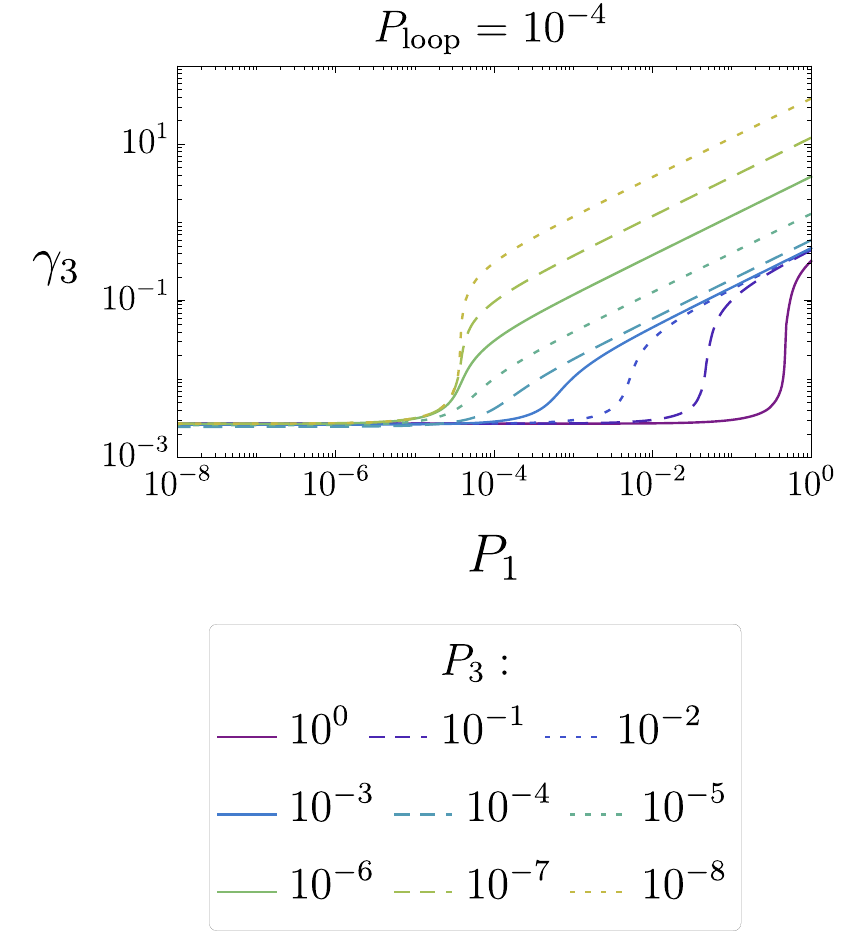}
  \end{minipage}
  \caption{Same as Fig.~\ref{plot-final value}, but as functions of $P_1 = P_2$ for $P_{\rm loop} = 10^{-4}$ in RD.
We take $P_3$ from $10^{0}$ to $10^{-8}$ (from top to bottom in the left panel, and from bottom to top in the right panel).
}
  \label{plot-final value vs P1}
\end{figure}

Figure~\ref{plot-final value vs P1} presents the same quantities as Fig.~\ref{plot-final value}, but plotted as functions of $P_1 = P_2$ for $P_{\rm loop} = 10^{-4}$ in the RD era.
Here, $P_3$ is varied from $10^{0}$ to $10^{-8}$ (from top to bottom in the left panel, and from bottom to top in the right panel).

The numerical results agree well with the analytical estimates obtained in the previous subsection.
From the left panels of Figs.~\ref{plot-final value} and~\ref{plot-final value vs P1},
$\gamma_1$ scales as $\propto P_3^{1/2}$ for $P_3 \gg P_{\rm loop}$ and becomes constant for $P_3 \ll P_{\rm loop}$ at fixed $P_{\rm loop}$,
independent of $P_1$, in agreement with Eq.~\eqref{gamma1 behavior for e}.

The dependence of $\gamma_3$ is more intricate but can also be confirmed from the right panels.
For very small $P_1$, $\gamma_3$ remains nearly constant at fixed $P_{\rm loop}$,
consistent with the first line of Eq.~\eqref{gamma3 behavior for e}.
The case $P_{\rm loop} \ll P_3 \ll P_1$ corresponds to the plateau regions in the right panel of Fig.~\ref{plot-final value}
and to the large-$P_1$ region for $P_3 \gtrsim 10^{-4}$ in Fig.~\ref{plot-final value vs P1}, showing $\gamma_3 \propto P_1^{1/2}$ as expected from the second line of Eq.~\eqref{gamma3 behavior for e}.
The final case, $P_3 \ll P_{\rm loop} \ll P_1$, exhibits a more complex dependence but is consistent with the third line of Eq.~\eqref{gamma3 behavior for e}.
This corresponds to the small-$P_3$ region with $P_1 \gtrsim 10^{-6}$ in Fig.~\ref{plot-final value}
and the large-$P_1$ region with $P_3 \lesssim 10^{-5}$ in Fig.~\ref{plot-final value vs P1}.

We also note 
that the scaling value of $\gamma_3$ changes abruptly
in the parameter region satisfying $P_{\rm loop} \ll P_3$ and $P_1 \simeq P_3$.
This behavior can be seen for the cases with $P_3 \sim P_1 = 10^{-1}, 10^{-2}, 10^{-3}, 10^{-4}$ in the right panel of Fig.~\ref{plot-final value} (where $P_{\rm loop} = 10^{-6}$) 
and for $P_1 \sim P_3 = 10^0, 10^{-1}, 10^{-2}$ 
in the right panel of Fig.~\ref{plot-final value vs P1} (where $P_{\rm loop} = 10^{-4}$). 
Moreover, a similar abrupt transition in the scaling value of $\gamma_3$ also appears in the parameter region satisfying
$P_3 \ll P_{\rm loop}$ and $P_1 \simeq P_{\rm loop}$. 
This corresponds to the cases with 
$P_1 \sim P_{\rm loop} = 10^{-4}$ for 
$P_3 = 10^{-8}, 10^{-7}, 10^{-6}$ 
in the right panel of Fig.~\ref{plot-final value vs P1}.

\begin{figure}
  \centering
   \includegraphics[width=0.7\linewidth]{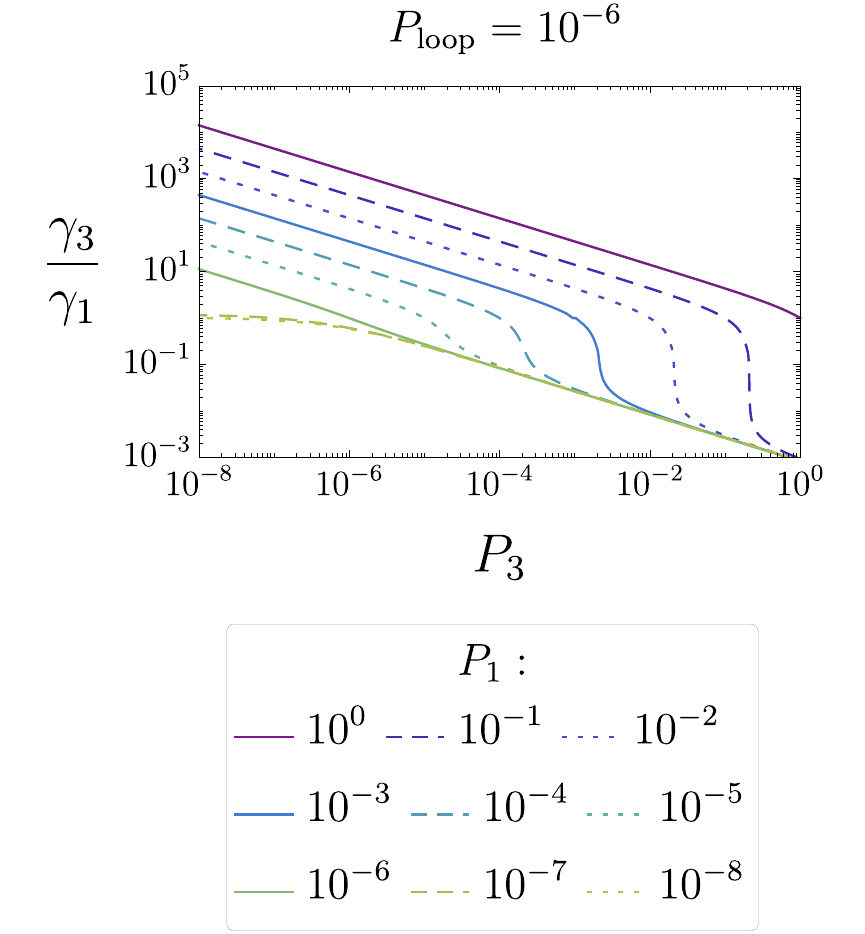}
  \caption{
   Scaling values of $\gamma_3/\gamma_1$ as a function of $P_3$ in the RD era,
for $P_{\rm loop} = 10^{-6}$.
We take $P_1 = P_2$ from $10^{0}$ to $10^{-8}$ (shown in different colors and line styles, from top to bottom in the right panel).
}
  \label{plot-ratio}
\end{figure}

Figure~\ref{plot-ratio} shows the ratio $\gamma_3/\gamma_1$ as a function of $P_3$.
This plot is useful for identifying which type of cosmic string dominates the energy density.
For example, if $\gamma_3/\gamma_1 \ll 1$, the component associated with $\gamma_3$ dominates the energy density. 
Typically, this occurs when $P_3 \gg P_1$ ($\gg P_{\rm loop}$), because in that case the probability of producing type-3 cosmic strings through the zippering process is relatively large.
Because we assume that all self-reconnection probabilities are identical, loops are predominantly produced by the dominant long-string component, and therefore this component also provides the main contribution to the GW signal.

If $P_{\rm loop} \gg P_i$ ($i = 1,2,3$), we expect $\gamma_3 / \gamma_1 \sim 1$ from Eqs.~(\ref{gamma1 behavior for e}) and (\ref{gamma3 behavior for e}).
This is because the evolution of the cosmic string network is dominated by the self-reconnection process, which is assumed to be identical for all string types.
This behavior is visible in Fig.~\ref{plot-ratio} at small values of $P_3$ for the cases with $P_1 = 10^{-7}$ and $10^{-8}$.
In contrast, for larger values of $P_3$, one finds that $\gamma_3/\gamma_1 \ll 1$ (or $\gg 1$) when $P_3 \gg P_1$ (or $P_1 \gg P_3$), respectively.

\subsection{Numerical results for the scaling time}
\label{sec:scalingtime}

Although one might expect the cosmic string network to reach the scaling regime shortly after formation,
the right panel of Fig.~\ref{plot-time evolution} shows that the approach to scaling can take a very long time,
depending on the reconnection probabilities.
We therefore determine numerically the characteristic time required for each string type to reach scaling.

We define the scaling time for type-$i$ strings ($i = 1$ or $3$) as the time when $\gamma_i$ reaches within $\pm 10\%$ of its scaling value.
We denote this as $t_{\text{scaling},i}$ ($i = 1,3$).

\begin{figure}
  \centering
  \begin{minipage}{0.48\linewidth}
  \includegraphics[width=\linewidth]{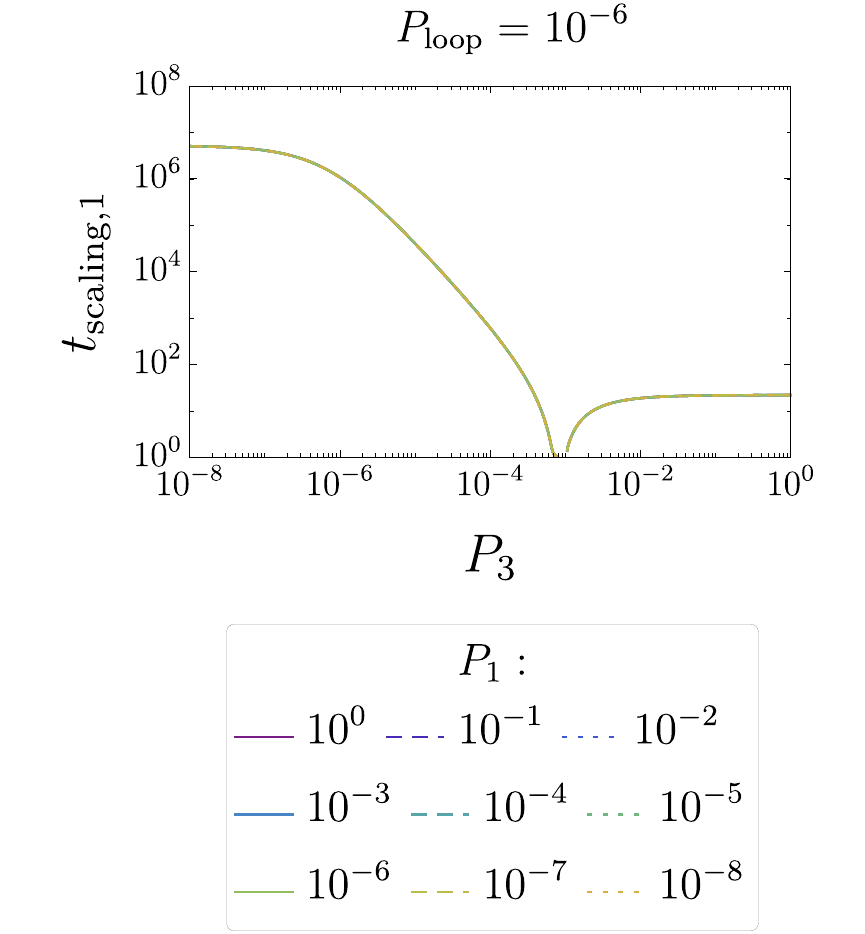}
  
  \end{minipage}
  \quad
  \begin{minipage}{0.48\linewidth}
  \includegraphics[width=\linewidth]{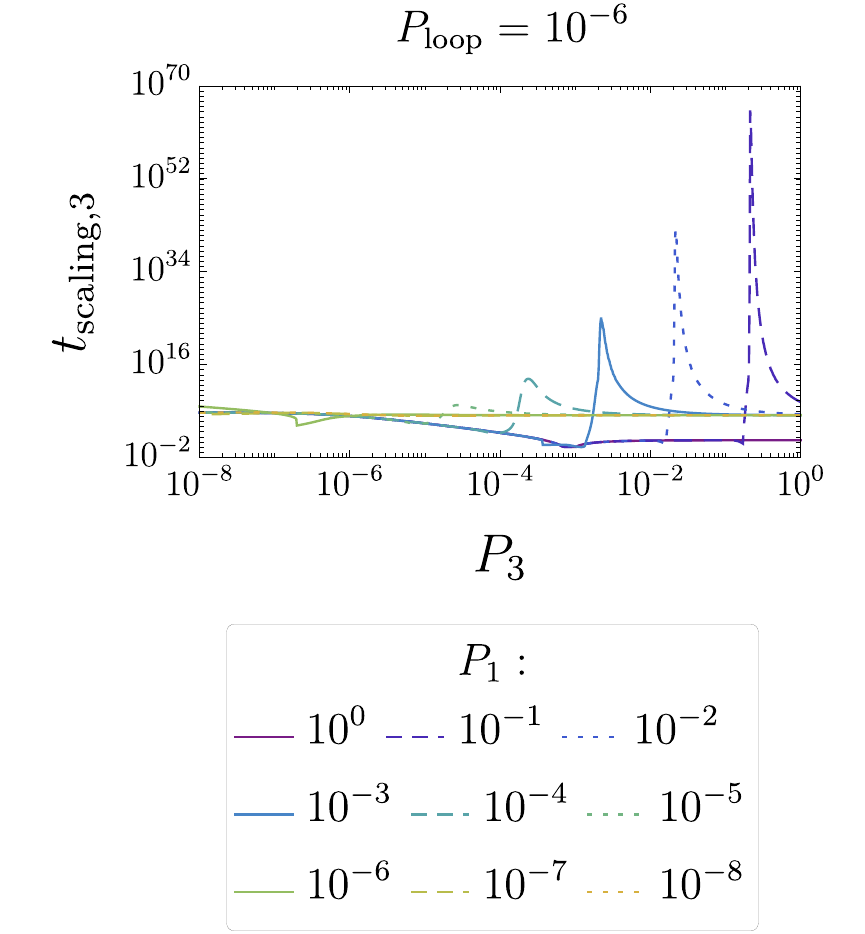}
  \end{minipage}
  \caption{
  Scaling times $t_{\text{scaling},1}$ (left) and $t_{\text{scaling},3}$ (right) as functions of $P_3$.
We vary $P_1$ from $10^{0}$ to $10^{-8}$, as indicated by different colors and line styles.
}
  \label{plot-scaling time}
\end{figure}

The left panel of Fig.~\ref{plot-scaling time} shows $t_{\text{scaling},1}$ as a function of $P_3$.
We vary $P_1$ from $10^{0}$ to $10^{-8}$, but all results are nearly degenerate.
$t_{\text{scaling},1}$ increases for moderately small $P_3$ and becomes independent of $P_3$ at very small $P_3$.
This can be understood from the initial condition $\gamma_1 = 10^{-2}$ and the scaling value given in Eq.~\eqref{gamma1 behavior for e}.
When the initial $\gamma_1$ is larger than its scaling value, corresponding to a lower initial string density, the reconnection process is initially inefficient,
and the evolution is dominated by Hubble expansion, which proceeds as a power law in time.
Consequently, it takes some time before the number density of long strings grows sufficiently for reconnections to establish scaling.
For the case of $P_3 \sim 10^{-3}$ and $P_{\rm loop} = 10^{-6}$,
the initial value of $\gamma_1$ is already close to its scaling value, and hence $t_{\text{scaling},1}$ is short.
For larger $P_3$, the initial string density exceeds the scaling value,
and the system reaches the rescaling regime relatively quickly via reconnections.

The right panel of Fig.~\ref{plot-scaling time} shows the behavior of $t_{\text{scaling},3}$.
For $P_1 = 10^{-1}$, $10^{-2}$, $10^{-3}$, and $10^{-4}$,
$t_{\text{scaling},3}$ exhibits a pronounced peak, increasing by several orders of magnitude around specific values of $P_3$ that depend on $P_1$.
These peaks occur near the threshold where the behavior of the scaling solution for $\gamma_3$ changes drastically, namely, in the region $P_{\rm loop} \ll P_3$ and $P_1 \simeq P_3$, as discussed above.
The peak value of $t_{\text{scaling},3}$ increases with larger $P_1$,
and we also confirm that it increases as $P_{\rm loop}$ decreases.
Moreover, we confirm that a similar enhancement of the scaling time $t_{\text{scaling},3}$ also appears in the parameter region satisfying
$P_3 \ll P_{\rm loop}$ and $P_1 \simeq P_{\rm loop}$,
which corresponds to the other regime in which the scaling value of $\gamma_3$ exhibits an abrupt transition, as discussed above.

\subsection{Physical origin of the ploronged scaling time}

The physical origin of the prolonged scaling time can be understood as follows.
First, we note from the numerical results that the scaling time for $\gamma_1$ is not significantly delayed.
Once $\gamma_1$ enters the scaling regime, it can be expressed in terms of $\gamma_3$ (see \eq{scaling eq of gamma3 for e}).
By substituting the scaling solution for $\gamma_1$, the VOS equation for $\gamma_3$ can be schematically written as
\begin{equation}
 \frac{d \ln \gamma_3}{d \ln t} = \sum_n a_n \gamma_3^n 
\end{equation}
where the coefficients $a_n$ are constants and 
$n$ takes integer values of $\mathcal{O}(1)$ 
(specifically $-2$, $0$, and $2$ in the extended VOS model, as seen from \eqref{eq:gamma3eq1} and \eqref{eq:gamma3eq2}).
In particular, $a_0$ contains $\mathcal{O}(1)$ numerical factors and therefore satisfies $\abs{a_0} \gtrsim \mathcal{O}(1)$ unless cancellations occur.

The scaling solution $\gamma_3 = x_s$ is determined by the condition $\sum_n a_n x_s^n = 0$.
Introducing a deviation from the scaling solution as $\gamma_3 = x_s (1 + \delta x)$,
the evolution equation for $\delta x$ becomes
\begin{equation}
 \frac{d \ln (1+\delta x)}{d \ln t} \sim \sum_n n a_n x_s^n \delta x\,,
\end{equation}
for small deviations, $\delta x \ll 1$. Using $\ln (1+\delta x) \simeq \delta x$, we obtain
\begin{equation}
 \frac{d \ln \delta x}{d \ln t} \sim \sum_n n a_n x_s^n. 
 \label{eq:deviationfromscaling}
\end{equation}
For a stable scaling solution, the right-hand side must be negative.

Since $\sum_n a_n x_s^n = 0$ and $\abs{a_0} \gtrsim \mathcal{O}(1)$, we generically expect 
\begin{equation}
 \abs{\sum_{n\ne 0 } a_n x_s^n} = \abs{a_0} \gtrsim  \mathcal{O}(1) \,.
\end{equation}
If no cancellations occur, one likewise expects
\begin{equation}
 \abs{\sum_{n\ne 0 } n a_n x_s^n} \gtrsim  \mathcal{O}(1) \,.
\end{equation}
so that the right-hand side of \eqref{eq:deviationfromscaling} remains of order unity or larger.
In this case, 
$\gamma_3(t)$ approaches the scaling regime via a power law, consistent with the naive expectation.
This behavior indeed appears when there are hierarchical differences among the reconnection probabilities.

However, the assumption
$\abs{\sum_{n\ne 0 } n a_n x_s^n} \gtrsim  \mathcal{O}(1)$
fails when the sum contains several terms of comparable magnitude that cancel among themselves.
This situation arises when $P_1 \sim {\rm Max}[ P_3, P_{\rm loop}]$. 
In such cases, the leading-order contribution is strongly suppressed, and the next-to-leading-order perturbations proportional to $(\delta x)^2$ become relevant.
As a result, the system approaches the scaling solution only logarithmically.

This explains why an extremely prolonged scaling time appears in a narrow region of parameter space, precisely where the above cancelation among the coefficients occurs.

From the discussion above, one can identify two essential ingredients in the VOS model that lead to a significantly delayed onset of the scaling regime.
The first is the presence of multiple interacting string species (three types in our case), which introduce several terms in the VOS equations with different dependences on the string densities.
The second is the existence of free parameters associated with the reconnection probabilities, which allows for the possibility of fine-tuning among these terms.
The combination of these two features enables the cancelations responsible for the delayed scaling behavior.
This phenomenon can therefore be regarded as a characteristic feature intrinsic to the string network considered in this work.
Since these features are also present in the conventional VOS model, they likewise lead to a significantly delayed onset of the scaling regime, as we explicitly confirm in Appendix~\ref{sec:Appendix}.

\section{GW signatures of delayed scaling cosmic string networks\label{sec:Gravitational Waves}}

According to our numerical results presented in Sec.~\ref{sec:RD}, the time required for the network to reach the scaling regime can increase by many orders of magnitude for multicomponent string systems with certain combinations of reconnection probabilities.
In such cases, the GW signals encode information about the network’s evolution prior to scaling, and the resulting spectrum is expected to exhibit nontrivial features.
In this section, based on the numerical solutions of the VOS equations, we compute the GW spectrum generated by the dynamics of the string network and discuss how it differs from the conventional scenario in which the network remains in the scaling regime throughout its evolution.

\subsection{Calculation method}

Here we briefly review the method used to calculate the GW spectrum from cosmic string loops, following Refs.~\cite{Caldwell:1991jj,DePies:2007bm,Sanidas:2012ee,Sousa:2013aaa,Sousa:2016ggw} (see also Ref.~\cite{Schmitz:2024gds}).

GWs are predominantly produced by string loops.
Since loop production proceeds through the self-reconnection process and is essentially independent of interactions among different types of cosmic strings, the resulting GW spectrum for each species can be computed independently once the time evolution of the long-string network is determined from the VOS equations.
The total GW spectrum is then obtained by summing the contributions from all string species.

We first recall that the effect of string loop production is phenomenologically described by the loop production function $f_i(l_p, t')$ that appears in \eqref{eq:looppf}.
We assume that it exhibits scaling behavior of the form
$f_i(l_p, t') = f_i(l_p/t') / t'$.
Motivated by numerical simulations, we then adopt a monochromatic form: 
\begin{align}
    f_i(x)=\frac{\mathcal{F}}{f_r}\frac{\tilde{c}_ic_{\xi_i}}{\alpha_i}\delta(x-\alpha_i).
    \label{eq:alpha}
\end{align}
The phenomenological parameters $\mathcal{F}$ and $f_r$ correct deviations caused by idealized assumptions.
The factor $\mathcal{F}$ corrects for the fact that the true distribution is not perfectly monochromatic~\cite{Sanidas:2012ee,Blanco-Pillado:2013qja}.
The factor $f_r$ accounts for the suppression of loop production due to nonzero center-of-mass energy, which decreases with redshift~\cite{Vilenkin:2000jqa}. 
We expect the typical loop size $\alpha_i$ to be of order unity, since the correlation length is of the order of the Hubble scale in the extended VOS model.
In our analysis we take
$f_r=\sqrt{2}$, $\mathcal{F}=0.1$, and $\alpha_i =0.1$~\cite{Blanco-Pillado:2013qja,Sousa:2016ggw,Auclair:2019wcv}.

After their formation, the loop length $l(t)$ decreases as the loops emit GWs.
Its evolution equation is
\begin{equation}
 \frac{d l}{dt} = - \Gamma G \mu_i \,,
\end{equation}
where $\Gamma\simeq 50$ is a numerical factor~\cite{Vachaspati:1984gt,Burden:1985md,Garfinkle:1987yw,Blanco-Pillado:2017oxo}. 
The solution is
\begin{equation}
 l(t) = l_p - \Gamma G \mu_i (t-t_p) \,,
 \label{eq:lt}
\end{equation}
where $l_p$ and $t_p$ denote the loop length and cosmic time at the moment of formation, respectively.
Under the assumption of \eqref{eq:alpha}, the loop length at a later time $t$ has a one-to-one correspondence with its initial length $l_p$.

Using \eqref{Energy~of~one~generated~loop} with $R \ll H^{-1}$ and $k(1/\sqrt2)$ for small loops, the evolution equation for the energy density of loops is given by
\begin{align}
    \dot\rho_{i,\text{loop}}(l_p,t)=-3H \rho_{i, \text{loop}}(l_p,t) + P_{\rm loop}\frac{\rho_i (t)v_i(t)l_p}{L_i^2}f_i(l_p,t)\,,\label{evolution eq for string loops}
\end{align}
where $\rho_{i, \text{loop}}(l_p,t)$ denotes the energy density of loops at time $t$. 
Here, we consider $\rho_{i, \text{loop}}$ as a function of $l_p$.
The first term represents the dilution due to Hubble expansion, while the second term represents the source term arising from chopping long strings via self-intersections.
Solving this equation formally yields
\begin{align}
    \rho_{i,{\rm loop}}(l_p,t)=P_{\rm loop}\mu_i l_p\int^t_{t_{\rm PT}}dt'\left(\frac{a(t')}{a(t)}\right)^3\frac{v_i(t')}{(\gamma_i(t')\,t')^4}f_i(l_p,t')\,,
    \label{eq:rholoop for gw}
\end{align}
where we use $L_i(t') = \gamma_i(t') t'$, and the lower limit of the integral corresponds to the onset of string formation at $t_{\rm PT}$.

The number density of loops of length $l$ at a time $t$ is 
\begin{align}
    n_{i, \text{loop}}(l,t)dl = \frac{\rho_{i, \text{loop}}(l_p,t)}{\mu_i l_p}dl_p\label{nuber density of loop} .
\end{align}
If $h_i(f,l)$ denotes the GW spectrum emitted by a loop of length $l$, the total GW energy emitted at time $t$ is 
\begin{align}
    \int dl ~n_{i, \text{loop}}(l,t)  h_i(f,l) \,.
\end{align}
By integrating over all physical frequencies $f$ and time from the onset of string formation ($t=t_{\rm PT}$) to the present, accounting for redshifting, the present-day GW energy density becomes
\begin{align}
    \rho_{\text{GW}}^i(t_0)=\int^{t_0}_{t_{\rm PT}}dt'\left(\frac{a(t')}{a(t)}\right)^3\int df\int_0^l dl~n_{i, \text{loop}}(l,t')h_i\left(\frac{a(t)}{a(t')}f,l\right)\,, \label{GW energy density}
\end{align}
where $t_0$ denotes the present time.

The average GW power emitted per loop, $h_i(f,l)$, is given by
\begin{align}
    h_i(f,l)=G\mu_i^2\sum^{\infty}_{n=1}P_n\delta(f-f_n(l))\,,
    \label{spectrum of GW}
\end{align}
with
\begin{align}
      P_n &= \frac{\Gamma}{\zeta(\frac{4}{3})}n^{-\frac{4}{3}}\,,\\
      f_n(l)&=\frac{2n}{l} \qquad ~(n =1,2,\dots)\,,
\end{align}
where cusp emission is assumed to be the dominant GW emission process. 
Here $n$ denotes the harmonic mode and $\zeta$ is the Riemann zeta function.
Using the delta functions, the $l$-integration in \eqref{GW energy density} can be performed explicitly, yielding
\begin{align}
    \frac{d\rho_\text{GW}^i}{d\ln f}&=G\mu_i^2f\sum_{n=1}^\infty P_n \frac{2n}{f^2}\int^{z_{\rm PT}}_0\frac{dz}{H(z)(1+z)^6}n_{i, \text{loop}}\left(\frac{2n}{(1+z)f},t'(z)\right) \,,
\end{align}
where 
$z = a(t_0)/a(t') -1$ denotes the redshift, and $z_{\rm PT}$ is its value at $t = t_{\rm PT}$.

Finally, by summing over all string species $i$,
the present-day GW spectrum is given by 
\begin{align}
    \Omega_{\text{GW}}(f)&\equiv \frac{8\pi G}{3H_0^2} \sum_i \frac{d\rho_{\text{GW}}^i}{d\ln f}\notag\\
    &=\frac{8\pi G^2}{3H_0^2} \sum_i \mu_i^2f \sum_{n=1}^{\infty}c_{n,i} (f)P_n\,,
    \label{GW density parameter}
\end{align}
with
\begin{align}
    c_{n,i}(f)=\frac{2n}{f^2}\int^{z_{\rm PT}}_0\frac{dz}{H(z)(1+z)^6}n_{i,\text{loop}} \left(\frac{2n}{(1+z)f},t'(z)\right)\,,
\end{align}
where $H_0$ is the Hubble parameter at present.

In computing the GW spectrum, Eqs.~\eqref{eveq of gamma with eff}, \eqref{eveq of v with eff}, \eqref{evolution eq for string loops}, and \eqref{GW density parameter} are solved numerically together with the Friedmann equation,
\begin{align}
\frac{1}{a}\frac{da}{dt} &= H_0 \left( \Omega_\Lambda + \Omega_m (1+z)^3 + \Omega_{\mathrm{rad}} \mathcal{G}(z)(1+z)^4 \right)^{1/2} \label{eq:Hubble}\\[6pt]
\mathcal{G}(z) &\equiv \frac{g_{*}(z)\, g_{s}^{4/3}(0)}{g_{*}(0)\, g_{s}^{4/3}(z)},
\end{align}
where $z$ is the redshift, $a(t)=1/(1+z)$ is the scale factor, $H_0 = 67.4 \, {\rm km/s/Mpc}$, $\Omega_\Lambda=0.685$, $\Omega_m=0.315$~\cite{Planck:2018vyg}, $\Omega_{\rm rad}=9.15 \times 10^{-5}$, and $g_*(z)$ and $g_s(z)$ are the effective numbers of relativistic degrees of freedom for energy and entropy density, respectively.

Before presenting the full numerical results, we summarize several important features of the GW spectrum that can be anticipated from the single-string-type analysis. 
Assuming RD and the scaling solution for the string density,
the integrals can be computed analytically, yielding
\begin{equation}
\lmk \Omega_{\rm GW} h^2 \rmk^{\rm (scaling, \, RD)} \propto \sum_i \Omega_{\rm rad} \sqrt{G\mu_i} \frac{P_{\rm loop}^i }{ \gamma_i^4} \,,
\label{eq:Omegascaling}
\end{equation}
in the extended VOS model, 
where numerical factors are omitted.
(See Ref.~\cite{Yamada:2022imq} for a detailed derivation.) 
Substituting the reconnection-probability dependence of $\gamma_i$ from \eqref{gamma1 behavior for e} and \eqref{gamma3 behavior for e},
we obtain the reconnection-probability dependence of the GW amplitude generated in the scaling regime:
\begin{align}
\lmk \Omega_{\rm GW} h^2 \rmk^{\rm (scaling, \, RD)} \propto 
\begin{cases}
P_{\text{loop}}^{-1} & \text{for } 
P_1 \ll {\rm Max}[P_3, \, P_{\text{loop}}] 
\text{ or } 
P_3 \ll P_\text{loop} \ll P_1 
\vspace{0.3cm}
\\
P_{\text{loop}} / P_3^2 & \text{for }
P_\text{loop} \ll P_3 \ll P_1 \\
\end{cases} \,. \label{eq:GWanalytic}
\end{align}
Therefore, when $(P_{\rm loop} / P_3) \ll 1$, the GW amplitude changes abruptly when $P_1$ is varied from $P_1 \ll P_3$ to $P_1 \gg P_3$. This threshold corresponds to the parameter region in which the scaling time becomes significantly delayed. This constitutes one of the main results of this paper.

In the realistic expansion history, the GW spectrum develops a peak at a characteristic frequency.
According to the full numerical calculations for the extended VOS model in Ref.~\cite{Yamada:2022imq}, the peak frequency scales as
\begin{equation}
\lmk f \rmk^{\rm (peak)} \propto (G \mu)^{-1} \,,
\label{eq:fpeak}
\end{equation}
and is independent of $P_{\rm loop}$ for the case of a single-type string.
We expect that this scaling also applies to our setup in the parameter region where the GW emission is dominated either by strings with tension $\mu_1$ (and $\mu_2$) or by those with tension $\mu_3$.

\subsection{GW spectrum with identical string tensions}

\begin{figure}
    \centering
    \begin{minipage}{0.47\linewidth}
        \centering
        \includegraphics[width=\linewidth]{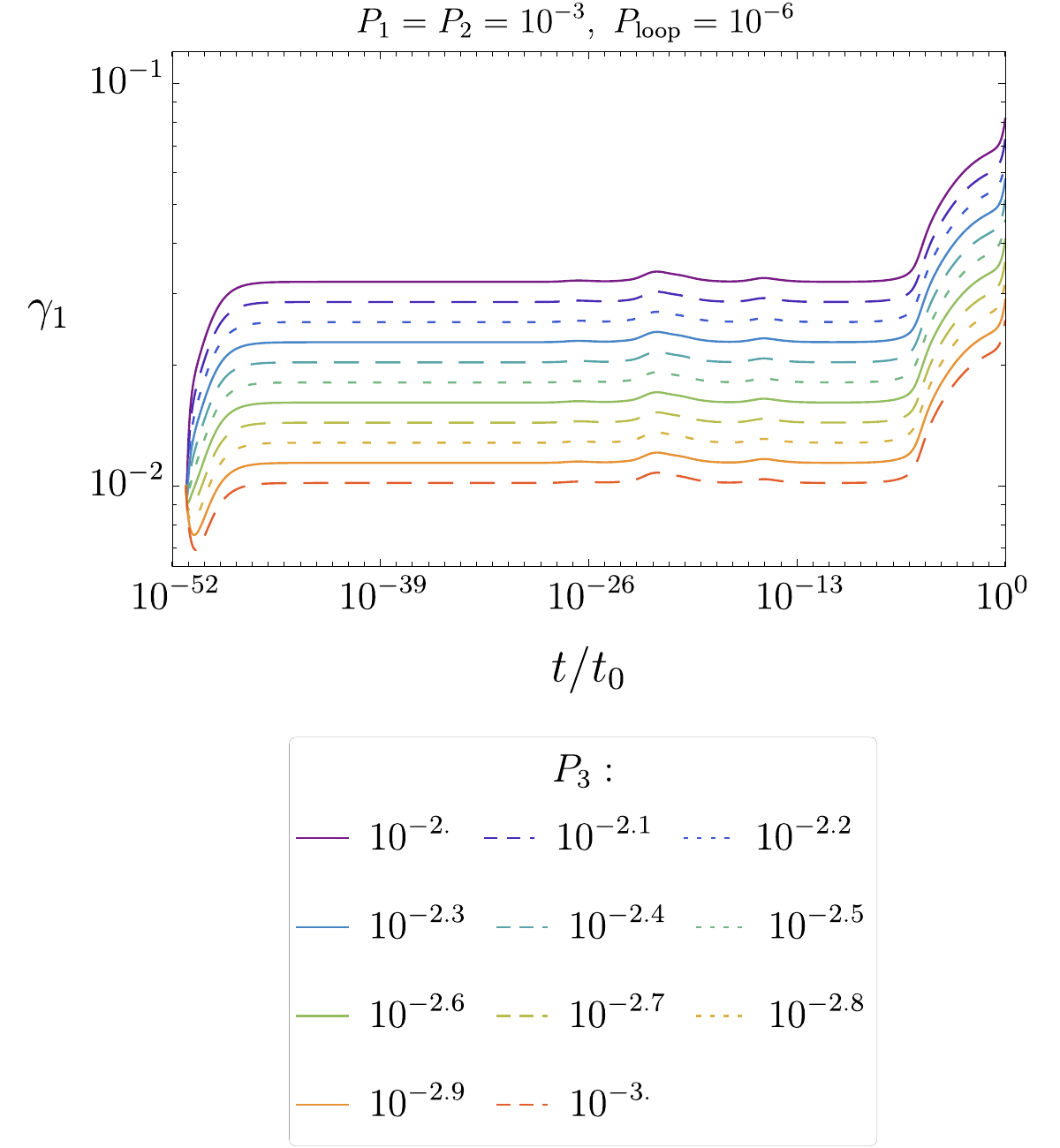}
    \end{minipage}
  \hspace{0.01\linewidth}
  \quad
    \begin{minipage}{0.47\linewidth}
        \centering
        \includegraphics[width=\linewidth]{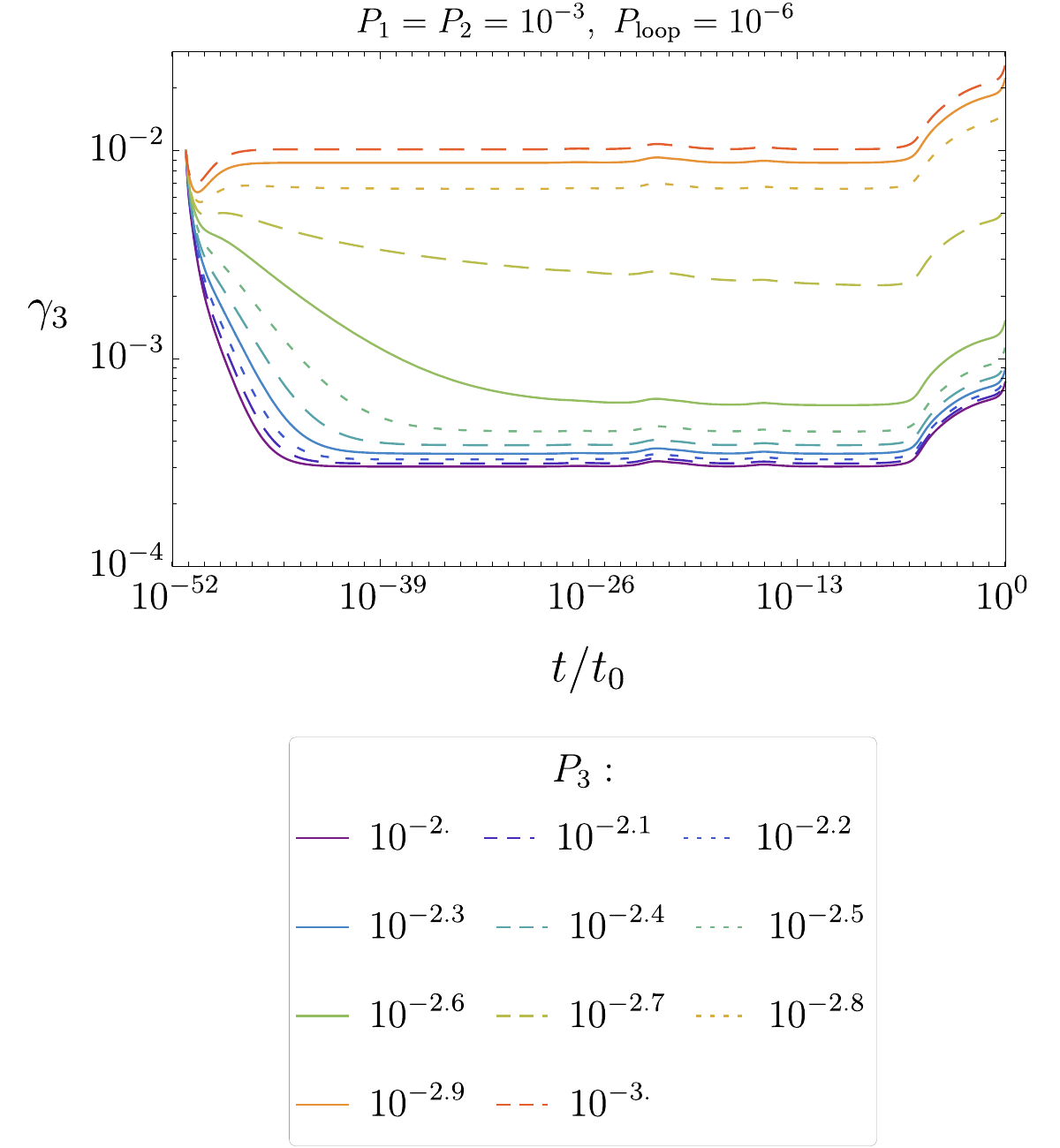}
    \end{minipage}
  \caption{Time evolution of $\gamma_i$ for the parameter set corresponding to Fig.~\ref{GW-delayscaling-byE}, including the effects of the realistic cosmic expansion history. The onset of the scaling regime is significantly delayed in some cases ($P_3 \sim P_1=P_2=10^{-3}$ and $P_\text{loop}=10^{-6}$, see also Fig.~\ref{plot-scaling time}). Each color corresponds to a different value of $P_3$.
  }
  \label{timeevoluton for GW-delayscaling-byE}
\end{figure}

We show numerical results of the GW spectrum for the case with $G\mu_i=10^{-12}$ for $i=1,2,3$.
Figure~\ref{timeevoluton for GW-delayscaling-byE} shows the time evolution of $\gamma_1=\gamma_2$ and $\gamma_3$ 
for a set of parameters chosen around those values that lead to a significantly delayed onset of the scaling regime.
We take the present age of the Universe, $t_0 \simeq 6.6 \times 10^{41} \GeV^{-1}$, as the time unit and set the initial time to $t_i/t_0 \simeq 7.2 \times 10^{-52}$ such that the initial temperature is of order $\sqrt{\mu_3}$ ($\simeq 1.2 \times 10^{13} \GeV$ for $G\mu_3 = 10^{-12}$). 
Since we adopt a realistic expansion history, starting from RD and transitioning to MD and finally to the dark-energy-dominated (DE) era, the behavior of $\gamma_i$ changes at the onset of MD (around $t/t_0 \sim 3.7 \times 10^{-6}$) even if the network has already reached the scaling regime during RD.
However, after the MD era, the evolution of $\gamma_i$ is similar for all parameter sets shown in Fig.~\ref{timeevoluton for GW-delayscaling-byE}.
We also emphasize that for certain parameter combinations, the time required to reach scaling in the RD era can be extremely long.

\begin{figure}
  \centering
  \vspace*{0.2cm}
  \includegraphics[width=0.9\linewidth]{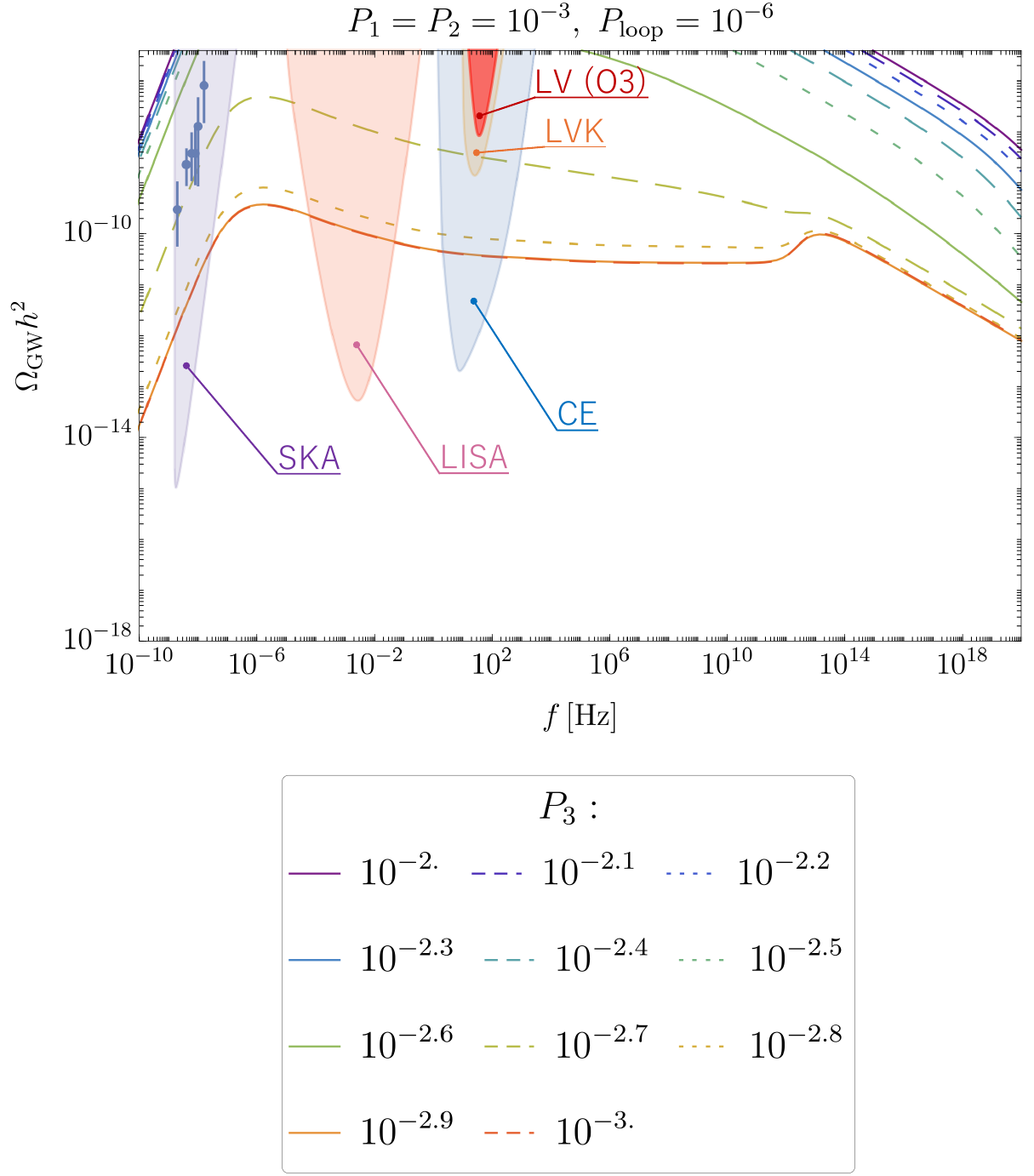}\\
  \caption{GW spectrum for a set of parameters chosen around those values that lead to a significantly delayed onset of the scaling regime. We take $G \mu_i = 10^{-12}$ ($i=1,2,3$), $P_1=P_2=10^{-3}$, and $P_\text{loop}=10^{-6}$, and vary $P_3$ from $10^{-2}$ to $10^{-3}$, as indicated by different colors and line styles. Pale shaded regions denote the projected sensitivities of future GW experiments, while the densely shaded red region is excluded by LIGO/Virgo. Data points with error bars correspond to the NANOGrav 15-year results~\cite{NANOGrav:2023gor}.}
  \label{GW-delayscaling-byE}
\end{figure}

Figure~\ref{GW-delayscaling-byE} shows the resulting GW spectra for the same parameter sets as in Fig.~\ref{timeevoluton for GW-delayscaling-byE}.
The high-frequency part of the spectrum is generated by small loops, whereas the low-frequency part mainly reflects the contribution from large loops.
If the network enters the scaling regime at late times, reconnections and loop production are less efficient, resulting in a reduced production of small loops during the early cosmological epochs.
Consequently, the high-frequency GW amplitude is suppressed.
Indeed, we observe that in parameter sets with delayed scaling, the GW signal at high frequencies is noticeably weaker.

Differences in the low-frequency part of the spectrum arise from variations in the scaling values of $\gamma_i$ for different reconnection probabilities $P_i$.
We have verified that, including the case for other parameter choices, the GW amplitude in the scaling regime agrees with the analytic behavior \eqref{eq:Omegascaling} expected from the corresponding $\gamma_i$.

\begin{figure}
  \centering
  \vspace*{0.2cm}
  \includegraphics[width=0.9\linewidth]{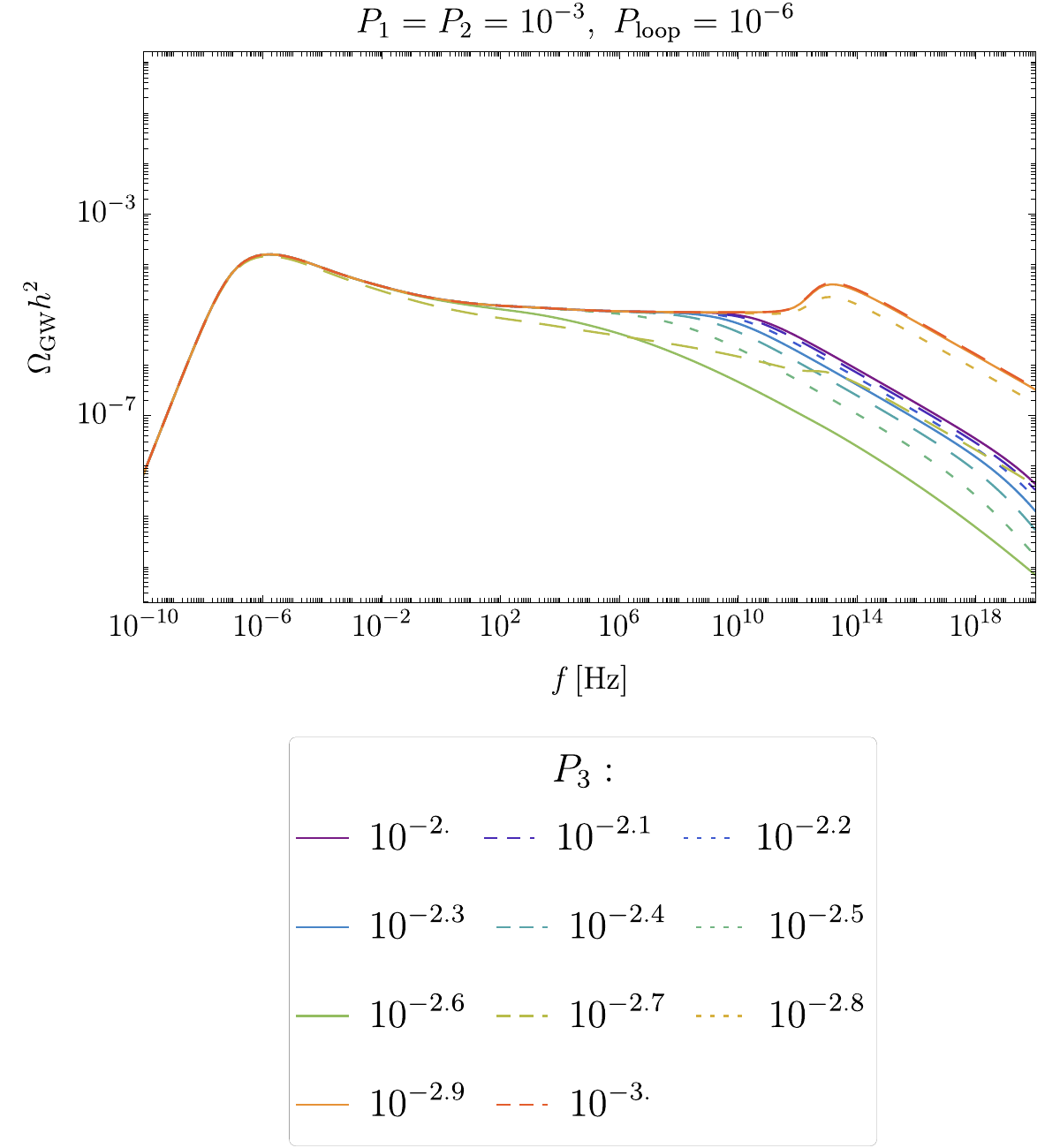}\\
  \vspace{0.7cm}
  \caption{Same as Fig.~\ref{GW-delayscaling-byE}, but with all spectra rescaled to coincide at low frequencies in order to highlight differences in the spectral shape.
  The overall normalization of the spectrum is not physically meaningful.}
  \label{freq dependence delayscaling-byE}
\end{figure}

The effect of delayed scaling is more clearly illustrated in Fig.~\ref{freq dependence delayscaling-byE}, where the GW amplitudes are rescaled to match in the low-frequency region.
If the network reaches the scaling regime sufficiently early, the spectra for all parameter sets exhibit similar amplitudes and shapes.
In contrast, when scaling is substantially delayed, the moderately high-frequency region is suppressed. 
The enhancement of the highest-frequency signals for the cases with $P_3 = 10^{-2.8}$, $10^{-2.9}$, and $10^{-3}$ may arise because the initial string density (or more precisely, the string density in the early regime shortly after the start of the simulation) can exceed the scaling value.
This corresponds to situations in which all initial values of $\gamma_i$ are smaller than their scaling values, making the GW amplitude sensitive to the initial conditions.

The GWs emitted by cosmic strings span a wide range of frequencies.
In Fig.~\ref{GW-delayscaling-byE}, 
the power-law-integrated sensitivity curves for ongoing and planned GW experiments 
are plotted according to Ref.~\cite{Schmitz:2020syl}, 
such as 
SKA~\cite{Janssen:2014dka},
LISA~\cite{LISA:2017pwj},
Cosmic Explorer (CE)~\cite{Reitze:2019iox},
and aLIGO+aVirgo+KAGRA (LVK)~\cite{Somiya:2011np,KAGRA:2020cvd}.
The red shaded region corresponds to the exclusion from the LV(O3) observing run~\cite{KAGRA:2021kbb}.
The data points with error bars correspond to the NANOGrav 15-year results~\cite{NANOGrav:2023gor}.
If the onset of the scaling regime is sufficiently delayed, the high-frequency slope becomes steeper.
This effect enlarges the viable parameter region that can simultaneously explain the NANOGrav signal while remaining consistent with LIGO/Virgo constraints.
Moreover, if a future GW detection reveals a similarly tilted high-frequency spectrum as predicted here, it would serve as compelling evidence for the delayed-scaling scenario.

\subsection{GW spectrum with hierarchical string tensions}

We also show numerical results for the GW spectrum in the case with a hierarchy of string tensions, $G\mu_1 = G\mu_2 \gg G\mu_3$.
This choice is motivated by the properties \eqref{D-string tension} and \eqref{F-string tension} of color flux tubes (or cosmic F- and D-strings) formed in $\mathrm{Spin}(4N)$ Yang-Mills theory for large $N$.

The time evolutions of $\gamma_1 = \gamma_2$ and $\gamma_3$ are the same as those in Fig.~\ref{timeevoluton for GW-delayscaling-byE},
since the evolution equation for long strings is independent of the string tensions.
The dependence on $\mu_i$ enters only through the time evolution of loop lengths via GW emission, \eqref{eq:lt}, and through the GW power emitted by loops, \eqref{spectrum of GW}.

\begin{figure}
  \centering
  \vspace*{0.2cm}
  \includegraphics[width=0.9\linewidth]{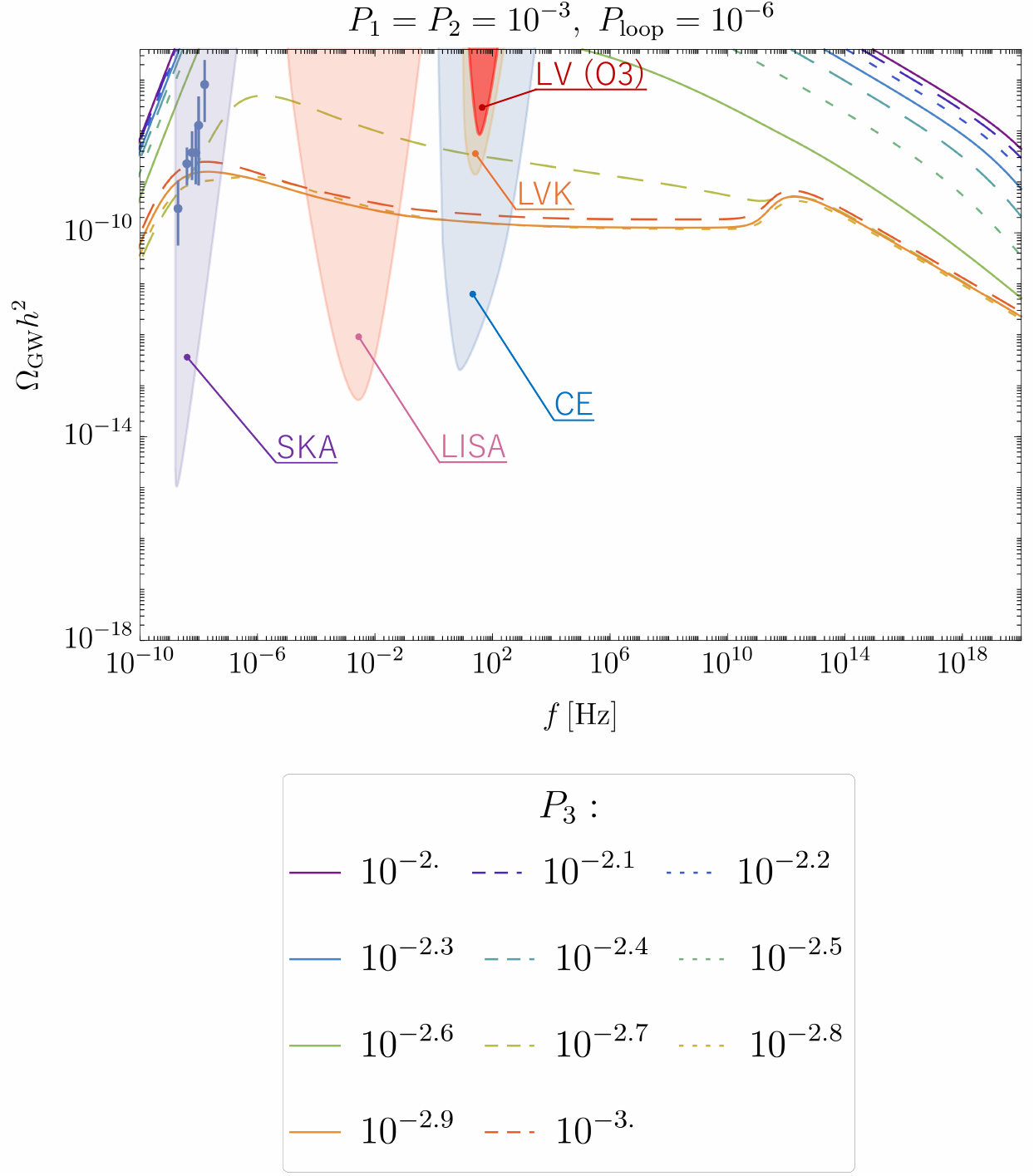}\\
  \caption{
  Same as Fig.~\ref{GW-delayscaling-byE} but with $G\mu_1=G\mu_2=10^2G\mu_3 = 10^{-10}$.
}
  \label{GW-delayscaling-byE-Gmu}
\end{figure}

To illustrate the effect of hierarchical string tensions, we specifically consider the case $G\mu_1 = G\mu_2 = 10^2 \, G\mu_3 = 10^{-10}$ as an example.
Figure~\ref{GW-delayscaling-byE-Gmu} shows the resulting GW spectra for this choice of string tensions.
Compared with the case in which all tensions are equal (Fig.~\ref{GW-delayscaling-byE}), no substantial change is observed for $P_3 \ge 10^{-2.7}$.
On the other hand, for $P_3 = 10^{-3} \,\text{-}\, 10^{-2.8}$, the peak frequency shifts to smaller values and the overall amplitude is enhanced.

To understand these features, let us parametrize the tension hierarchy as
\begin{align}
    \mu_1=\mu_2=N\mu_3\label{hierarchy of tension}\,,
\end{align}
and consider the case with $N \gg 1$.
We first note that the nearly flat plateau in the GW spectrum originates from GWs emitted by the scaling string network during RD. 
In this regime, \eqref{GW density parameter} implies
\begin{align}
   \lmk\Omega_{\rm{GW}}h^2\rmk&\propto \sqrt{G\mu_3}
   \lmk
   \sqrt{N}\frac{P_\text{loop}}{\gamma_1^4}+\sqrt{N}\frac{P_\text{loop}}{\gamma_2^4}+\frac{P_\text{loop}}{\gamma_3^4}
   \rmk\notag\\[5pt]
   &=P_\text{loop}\frac{\sqrt{G\mu_3}}{\gamma_3^4}\lmk
   2\sqrt{N}\lmk\frac{\gamma_3}{\gamma_1}\rmk^4+1\rmk
\end{align}
where we used the fact that $\gamma_1=\gamma_2$.
Therefore, the amplitude does not change significantly if
$\gamma_3 / \gamma_1 \ll N^{-1/8}$, while it scales as $N^{1/2}$ otherwise. 
From \eqref{ratio behavior for e}, the condition $\gamma_3 / \gamma_1 \ll N^{-1/8}$ can be satisfied when
$P_1, {N}^{1/4}\, P_{\rm loop} \ll P_3$. 
This marginally corresponds to $P_3 \ge 10^{-2.7}$ in the example shown in Fig.~\ref{GW-delayscaling-byE-Gmu}, for which no significant change is observed even for $N = 10^2$.
Conversely, when $P_3 < 10^{-2.7}$, the plateau amplitude is enhanced by a factor of $\sim \sqrt{N} = 10$.
Moreover, 
as noted in \eqref{eq:fpeak}, the peak frequency scales as the inverse of the string tension, explaining why the peak shifts to a lower frequency in the $P_3 < 10^{-2.7}$ cases shown in Fig.~\ref{GW-delayscaling-byE-Gmu}.
We have checked that a similar behavior is realized for different values of $N$, in agreement with the analytic arguments above.

\section{Summary and discussion\label{sec:Summary and discussion}}
We have analyzed the dynamics of cosmic string networks composed of three types of strings using the extended VOS model, taking into account reconnections among the strings with small probabilities.
This setup is motivated by a pure Spin($4N$) gauge theory, which predicts three kinds of macroscopic color flux tubes formed after the confinement phase transition.
We analytically derived the dependence of the scaling solution on the reconnection probabilities and confirmed these results through numerical simulations.
The scaling solutions provide valuable insight into which string species dominate the energy density of the cosmic string network and how efficiently GWs are emitted from loops of the dominant species through self-reconnection.
Furthermore, we computed the GW spectrum emitted by closed loops of cosmic strings, incorporating the cosmic expansion history.

Our analysis revealed that there exist parameter regions in which the string network requires a very long time to reach the scaling regime. 
We find that this behavior is a distinctive feature of systems with multiple string species and interactions characterized by small reconnection probabilities.
In such cases, the GW amplitude is suppressed in the high-frequency region because loop production is inefficient during the early epoch.
This feature allows the model to better accommodate the NANOGrav signal while remaining consistent with current LIGO/Virgo constraints, although the original fast-scaling scenario already provides a good fit to the PTA data. 
The scenario can be further tested with future GW observations, such as CE and LISA, which are sensitive to higher-frequency bands beyond the nanohertz range.
Since the predicted GW spectra span a wide range of frequencies and the differences among parameter sets become more pronounced at higher frequencies, a future detection of a high-frequency spectral slope consistent with our predictions would provide strong support for this delayed-scaling scenario.

Our results can also be relevant to cosmic superstring networks predicted in D-brane inflation models.
In such models, the relations between string tensions and reconnection probabilities have been studied in detail, and the reconnection probability is typically estimated to be of order $10^{-(3\,\text{-}\,2)}$ or larger~\cite{Jackson:2004zg,Hanany:2005bc}.
Moreover, additional effects such as volume suppression in the compact dimensions~\cite{Jones:2003da,Jackson:2004zg,Pourtsidou:2010gu} can lead to even smaller effective reconnection probabilities.
Therefore, the parameter space explored in our analysis is also plausible in this context.
Although cosmic superstring scenarios can in principle involve an infinite number of string species, previous studies have shown that the network evolution is often dominated by only the first few lightest species~\cite{Pourtsidou:2010gu}.
Thus, our analysis of a network composed of three distinct string species can serve as a useful effective description in such cases.

GWs are also predicted from the dynamics of the confinement/deconfinement phase transition~\cite{Reichert:2021cvs,Morgante:2022zvc,He:2022amv,Reichert:2022naa,Pasechnik:2023hwv}, although the peak frequency is much higher than the frequency range relevant for the parameter space considered in this work.
These signals are complementary to those from cosmic strings, providing two distinct sources of GWs associated with the confinement/deconfinement phase transition: one from the string network and the other from a first-order phase transition.

In the present paper, we focus on the macroscopic color flux tubes that are formed during the confinement/deconfinement phase transition.
However, it is important to note that glueballs are also produced simultaneously during the phase transition.
In most gauge theories, these glueballs are unstable and decay into Standard Model particles through higher-dimension operators, and therefore they are not viable dark matter candidates in the parameter region of our interest.%
\footnote{
See also Ref.~\cite{Faraggi:2000pv,Feng:2011ik,Boddy:2014yra,Boddy:2014qxa,Soni:2016gzf,Kribs:2016cew,Forestell:2016qhc,Soni:2017nlm,Forestell:2017wov,Jo:2020ggs,Carenza:2022pjd,Carenza:2023shd,Bruno:2024dha,Biondini:2024cpf,Carenza:2024avj} for scenarios of light glueball dark matter, where the confinement scale is much smaller than in our setup.
}
In contrast, in $\mathrm{SO}(2N)$ or Spin($2N$) gauge theories, including the model considered in this paper, a special glueball state, referred to as the baryonic glueball, can acquire a very long lifetime due to an accidental symmetry inherent in the gauge structure, making it a viable dark matter candidate~\cite{Gross:2020zam,Yamada:2023thl}.
Our recent work further suggests that its relic abundance can be consistent with the observed dark matter density for parameter values of interest, specifically around $G\mu \sim 10^{-12}$~\cite{Yamada:2023thl}.

\appendix

\section{Analytic calculation of scaling solution
\label{sec:Analytic behavior of scaling solution in Extended VOS}}

In this Appendix, we present the detailed derivation of the analytic scaling solution.
For simplicity, we focus on the RD era ($H = 1/2t$).

We solve $\dot{\gamma}_i = \dot{v}_i = 0$ in \eq{eveq of gamma with eff} and \eq{eveq of v with eff}, assuming $B =0$.
Equivalently, the evolution equation for the velocity reduces to
\begin{align}
    \dot{v}_i&=\frac{1-v_i^2}{t}\Big(\frac{k(v_i)}{c_\xi}-v_i\Big)\label{eveq of v}.
\end{align}
The scaling solution for the velocity therefore satisfies
\begin{align}
    v_i = \frac{k(v_i)}{c_{\xi}} \,,
\end{align}
and all components share the same value,
\begin{align}
    v_i= v = \frac{k(v)}{c_{\xi}} \quad ({\text{for any }}i)\,,
\end{align}
implying that the velocity approaches a constant in the scaling regime.

Next, we simplify the expressions by omitting subscripts and constant numerical factors, rewriting the effective reconnection probabilities as
\begin{equation}
\begin{aligned}
    \tilde{c}c_\xi vP_{\rm loop} &\to P_{\rm loop}\,, \notag\\
    d\bar{v}c_{\xi}P_1 &\to P_{\text{1}}\,, \notag\\
    d\bar{v}c_{\xi}P_2 &\to P_{\text{2}}\,, \notag\\
    d\bar{v}c_{\xi}P_3 &\to P_{\text{3}}\,. \notag 
\end{aligned}
\end{equation}
Since $v$ approaches a constant, its explicit dependence can be safely absorbed into these redefinitions.

\subsection{Behavior of $\gamma_1$}

In the scaling regime, $\dot{\gamma}_1=0$, and substituting Eqs.~\eqref{simplification1 for e}-\eqref{simplification3 for e} into \eqref{eveq of gamma with eff} (using also $\gamma_1=\gamma_2$ and $P_1 = P_2$), we obtain
\begin{align}
    0=(1+v^2)-2&+\frac{P_{\text{loop}}}{\gamma_1^2}+\frac{P_3}{\gamma_1^2}.\label{scaling eq of gamma1 for e}
\end{align}
This is a quadratic equation for $\gamma_1$, and approximating $(1+v^2)-2\simeq -1$ yields
\begin{align}                   
    \gamma_1^2&=P_{\rm{loop}}+{P_3}\notag\\
    \gamma_1 &=\sqrt{P_{\rm{loop}}+P_{3}} \,.
\end{align}
Thus the asymptotic scaling of $\gamma_1$ is
\begin{align}
\gamma_1 &=\mathcal{O}(1)\times
\begin{cases}
\sqrt{P_{\text{loop}}} & \text{for } P_3 \ll P_{\text{loop}} \vspace{0.3cm}
\\
\sqrt{P_3} & \text{for }
P_{\text{loop}} \ll P_3 \,.
\end{cases} 
\label{gamma1 behavior in extended vos}
\end{align}

\subsection{Behavior of $\gamma_3$ }
In the same manner as for $\gamma_1$, we obtain the following equation that the scaling solution for $\gamma_3$ should satisfy:
\begin{align}
    0=(1+v^2)-2+\frac{P_{\text{loop}}}{\gamma_3^2}+2\frac{P_1}{\gamma_1^2}
    -\frac{\gamma_3^2}{\gamma_1^4}P_3\label{scaling eq of gamma3 for e}.
\end{align}
This equation is more complicated than \eqref{scaling eq of gamma1 for e} because it contains both $\gamma_1$ and $\gamma_3$.
Therefore, we analyze different parameter regimes separately.
In what follows, we approximate $(1+v^2)-2\approx -1$.

\subsubsection{$P_3 \ll P_\text{loop}$}
In this regime, we have $\gamma_1^2 \approx P_\text{loop}$ from \eqref{gamma1 behavior in extended vos}.
Substituting this into \eqref{scaling eq of gamma3 for e}, we obtain
\begin{align}
    0=-1+\frac{P_{\text{loop}}}{\gamma_3^2}+2\frac{P_1}{P_{\text{loop}}}
    -\frac{\gamma_3^2}{P_{\text{loop}}^2}P_3 \,,
    \label{eq:gamma3eq1}
\end{align}
which reduces to the quartic equation
\begin{align}
    \frac{P_3}{P_{\text{loop}}^2}\gamma_3^4+\Big(1-\frac{2P_1}{P_\text{loop}}\Big)\gamma^2_3-P_\text{loop}=0.
\end{align}
The solution is given by
\begin{align}
    \gamma^2_3 = \frac{P_\text{loop}^2}{2P_3}\left\{-\left(1-\frac{2P_1}{P_\text{loop}}\right) + \sqrt{\left(1-\frac{2P_1}{P_\text{loop}}\right)^2+\frac{4P_3}{P_\text{loop}}}\right\}\,,
\end{align}
where a negative root is discarded because $\gamma_3^2$ must be positive.

If $P_3\ll P_\text{loop}\ll P_1$, the solution is approximately
\begin{align}
    \gamma_3^2
    &\approx 2\frac{P_\text{loop}P_1}{P_3} \,.
\end{align}
Thus $\gamma_3$ scales as $P_1^{1/2}P_3^{-1/2}P_\text{loop}^{1/2}$.

If instead $P_1\ll P_3 \ll P_\text{loop}$ or $P_3\ll P_1 \ll P_\text{loop}$, we obtain 
\begin{align}
    \gamma_3^2&\approx
\frac{P_\text{loop}^2}{2P_3}\left\{-\left(1-\frac{2P_1}{P_\text{loop}}\right)+\left(1-\frac{2P_1}{P_\text{loop}}\right)+\frac{4P_3}{2P_\text{loop}}\right\}\notag\\
&=P_{\text{loop}} \,.
\end{align}
In this regime, $\gamma_3$ is proportional to $P_\text{loop}^{1/2}$.
The other root is discarded because $\gamma_3$ must be positive.

\subsubsection{$P_\text{loop}\ll P_{3}$}

Here, we use $\gamma_1^2 \approx{P_3}$ from \eqref{gamma1 behavior in extended vos}.
Proceeding as before, we obtain the quartic equation
\begin{align}
    \frac{1}{P_3}\gamma_3^4+\left(1-\frac{2P_1}{P_3}\right)\gamma_3^2-P_\text{loop}=0.
    \label{eq:gamma3eq2}
\end{align}
Its solution is
\begin{align}
    \gamma_3^2 =\frac{P_3}{2}\left\{-\left(1-\frac{2P_1}{P_3}\right) + \sqrt{\left(1-\frac{2P_1}{P_3}\right)^2+\frac{4P_\text{loop}}{P_3}}\right\}\,,
\end{align}
where a negative root is discarded because $\gamma_3^2$ must be positive.

If $P_\text{loop}\ll P_3\ll P_1$, we obtain 
\begin{align}
    \gamma_3^2 
    &\approx 2 P_1 \,.
\end{align}
In this case, $\gamma_3$ is proportional to $P_1^{1/2}$.

If $P_\text{loop}\ll P_1 \ll P_3$ or $P_1\ll P_\text{loop} \ll P_3$, we obtain 
\begin{align}
    \gamma_3^2&\approx
\frac{P_3}{2}\left\{-\left(1-\frac{2P_1}{P_3}\right)+\left(1-\frac{2P_1}{P_\text{loop}}\right)+\frac{4P_\text{loop}}{2P_3}\right\}\notag\\
&=P_{\text{loop}} \,.
\end{align}
and thus $\gamma_3 \propto P_\text{loop}^{1/2}$.
Again, we discard the negative root.

\subsection{Summary}

The scaling behavior of $\gamma_1$ and $\gamma_3$ is summarized as follows.
\begin{align}
\gamma_1 &=\mathcal{O}(1)\times
\begin{cases}
\sqrt{P_{\text{loop}}} & \text{for } P_3 \ll P_{\text{loop}} \vspace{0.3cm}
\\
\sqrt{P_3} & \text{for }
P_{\text{loop}} \ll P_3
\end{cases}
\\
\notag\\
\gamma_3 &=\mathcal{O}(1)\times
\begin{cases}
\sqrt{P_{\text{loop}}} & \text{for } 
P_1 \ll {\rm Max}[P_3, \, P_{\text{loop}}] \vspace{0.3cm}
\\
\sqrt{P_1}& \text{for }
P_\text{loop} \ll P_3 \ll  P_1 \vspace{0.3cm}
\\
P_1^\frac{1}{2}P_3^{-\frac{1}{2}}P_\text{loop}^{\frac{1}{2}} & \text{for }
P_3 \ll P_\text{loop}\ll P_1 \,.\\
\end{cases}
\end{align}

\section{Conventional VOS model}
\label{sec:Appendix}

In this Appendix, we consider the conventional VOS model for multiple string species with small reconnection probabilities.

In the literature, it is often assumed that small reconnection probabilities can be incorporated by effectively replacing $P_{\rm loop}^i \to (P_{\rm loop}^i)^{1/3}$ and $P_i \to P_i^{1/3}$ in the VOS equations.
Moreover, the correlation length $\xi_i$ and the inter-string distance $L_i$ are assumed to be equal, namely
\begin{equation}
 \xi_i = L_i\,,
\end{equation}
instead of \eq{eq:xi}.
Correspondingly, the typical zipper length $\ell^i_{jk}(t)$ is written in terms of $L_i$ ($i=1,2,3$) as
\begin{align}
    \ell^i_{jk}(t)&=\frac{L_j L_k}{L_j+L_k}=\frac{\gamma_j \gamma_k}{\gamma_j+\gamma_k}t\,,
    \label{simplification3}
\end{align}
where we define 
\begin{equation}
 L_i(t)=\gamma_i(t)t\,.
\end{equation}
The typical loop size is expected to be of the order of the correlation length, which is assumed to be equal to the inter-string distance in the conventional VOS model.
Consequently, the parameter $\alpha_i$ in \eq{eq:alpha} is taken to be $\alpha_i = 0.3 P_{\rm loop}^{1/3}$, which corresponds to the scaling solution of $\gamma_i$ in the case without interactions.%
\footnote{
Although the scaling solution differs in particular when $P_{\rm loop} \ll P_i$,
we adopt the conventional value in this Appendix.
}

Then, the VOS equations become
\begin{align}
    \frac{\dot{\gamma}_i}{\gamma_i}&=\frac{1}{2t}\Bigg(2Ht(1+v_i^2)-2+(P_{\rm loop}^i)^{1/3}\frac{\tilde{c}_iv_i}{\gamma_i}+P_k^{1/3}\tilde{d}^k_{ij}\bar{v}_{ij}\frac{1}{\gamma_j^2}\frac{\ell^k_{ij}(t)}{t}\notag
    \\&~~~~+P_j^{1/3}\tilde{d}^j_{ki}\bar{v}_{ki}\frac{1}{\gamma_k^2}\frac{\ell^j_{ki}(t)}{t}-P_i^{1/3}\tilde{d}^i_{jk}\bar{v}_{jk}\frac{\gamma_i^2}{\gamma_j^2\gamma_k^2}\frac{\ell^i_{jk}(t)}{t}\Bigg)\label{seveq for gamma}\\[5pt]
    \dot{v}_i&=\frac{1-v_i^2}{t}\left(\frac{k(v_i)}{\gamma_i}-2Htv_i+ B P_i^{1/3}\tilde{d}^i_{jk}\frac{\bar{v}_{jk}}{v_i}\frac{\mu_j+\mu_k-\mu_i}{\mu_i}\frac{\gamma_i^2}{\gamma_j^2\gamma_k^2}\frac{\ell^i_{jk}(t)}{t}\right).\label{seveq for v}
\end{align}

Moreover, we are interested in cases where the type-1 and type-2 strings behave identically due to symmetry.
Thus, we impose $\mu_1 = \mu_2$, $P_1 = P_2$, and $\gamma_1 = \gamma_2$.

We further simplify 
$\mathcal{O}(1)$ numerical factors as well as parameters 
by assuming Eqs.~\eqref{eq:tildec}, \eqref{simplification2 for e}, \eqref{simplification1 for e}, \eqref{simplification3 for e}, and \eqref{eq:Ploopi}. 
In our numerical calculations, we take $\tilde{c} = \tilde{d} = 0.23$ and choose $c_\xi$ according to \eq{eq:cxi}.

\subsection{Analytic behavior of the scaling solution}
\label{sec:B1}
Here we show the analytic behavior of the scaling solution in the conventional VOS model.

The scaling behavior of the velocity can be understood as follows.
For simplicity, we take $B=0$ in \eqref{seveq for v}.
Then the scaling solution satisfies
\begin{align}
    v_i =\frac{k(v_i)}{\gamma_i} \,.
    \label{scaling relation of v}
\end{align}
As will be shown later, $\gamma_i$ approaches a scaling regime in which its value is proportional to some combination of reconnection probabilities and becomes very small when the reconnection probabilities are small.
Thus the factor $1/\gamma_i$ in \eqref{scaling relation of v} can become very large.
However, the left-hand side is bounded above by unity.
This implies that the momentum parameter $k(v_i)$ must be small, and consequently $v_i$ must approach $1/\sqrt{2}$ since $k(1/\sqrt{2})=0$.
Therefore, we may treat $v_i$ as approximately constant when deriving the analytic scaling behavior of $\gamma_i$.

The analytic calculation of the scaling solution for $\gamma_i$ is given in Appendix~\ref{sec:analyticforconventional}.
Here we summarize the results:
\begin{align}
\label{eq:gamma1scalingcon1}
\gamma_1 &=\mathcal{O}(1)\times
\begin{cases}
P_{\rm loop}^{1/3} & \text{for } P_3\ll P_{\rm loop} \vspace{0.3cm}
\\
P_3^{1/3} & \text{for }
P_{\rm loop} \ll P_3
\end{cases}
\\
\notag\\
    \gamma_3 &=\mathcal{O}(1)\times
\begin{cases}
P_{\rm loop}^{1/3} & \text{for }\quad
P_1 \ll {\rm Max}[ P_{\rm loop}, \,  P_3  ]
\vspace{0.3cm}\\
P_1^{1/6} P_3^{1/6}& \text{for}\quad
P_{\rm loop} \ll P_3 \ll  P_1 
\vspace{0.3cm}\\
P_1^{1/6}P_3^{-1/6}P_{\rm loop}^{1/3} & \text{for}\quad
P_3 \ll P_{\rm loop}\ll P_1\,, \\
\end{cases}
\label{eq:gamma1scalingcon2}
\end{align}
and $\gamma_2 = \gamma_1$, where the $\mathcal{O}(1)$ prefactors include the parameters 
$\tilde{c}$, $c_\xi$, $\tilde{d}$, and $v$.

\subsection{Numerical results}

We performed numerical simulations for the conventional VOS model using Eqs.~\eqref{seveq for gamma} and \eqref{seveq for v}.
The initial values of the variables were chosen as
\begin{align}
    \gamma_i=10^{-2},\, \qquad v_i=10^{-2}\qquad\text{for }\, i=1,2,3 \,.
\end{align}

\begin{figure}
  \centering
  \begin{minipage}{0.49\linewidth}
  \includegraphics[width=\linewidth]{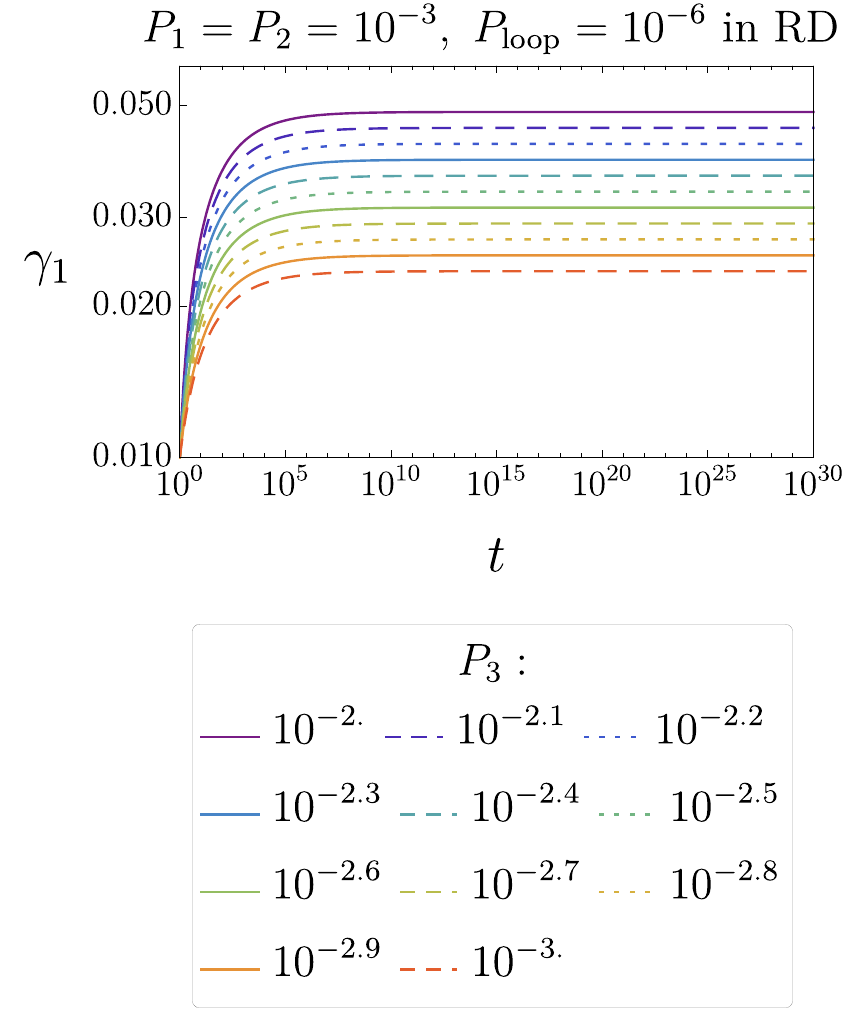}
  \end{minipage}
  \begin{minipage}{0.49\linewidth}
  \includegraphics[width=\linewidth]{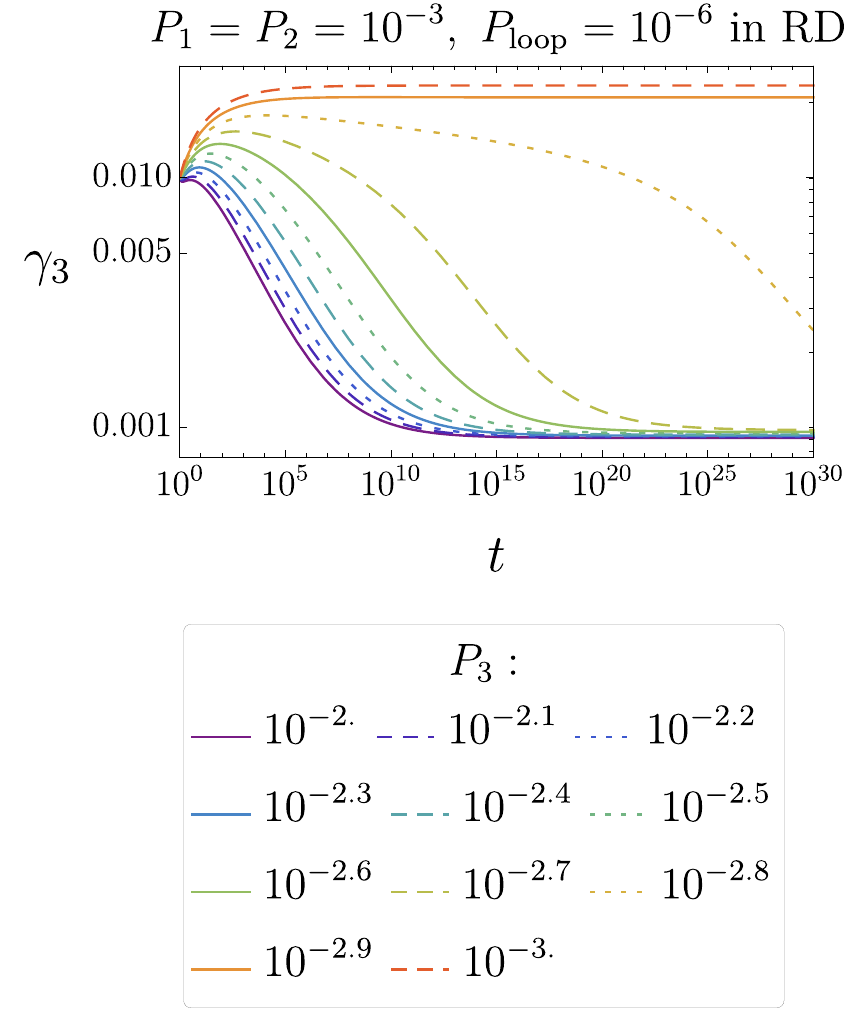}
  \end{minipage}
  \caption{Same as Fig.~\ref{plot-time evolution} but for the conventional VOS model.}
  \label{plot-time evolution forS}
\end{figure}

We first consider the case of the RD era.
Figure~\ref{plot-time evolution forS} shows the time evolution of $\gamma_1 = \gamma_2$ (left panel) and $\gamma_3$ (right panel) for $P_1 = P_2 = 10^{-3}$ and $P_{\rm loop} = 10^{-6}$, with several values of $P_3$.
Each $\gamma_i$ asymptotically approaches a constant value and reaches a scaling regime at late times.

\begin{figure}
  \centering
  \begin{minipage}{0.49\linewidth}
  \includegraphics[width=\linewidth]{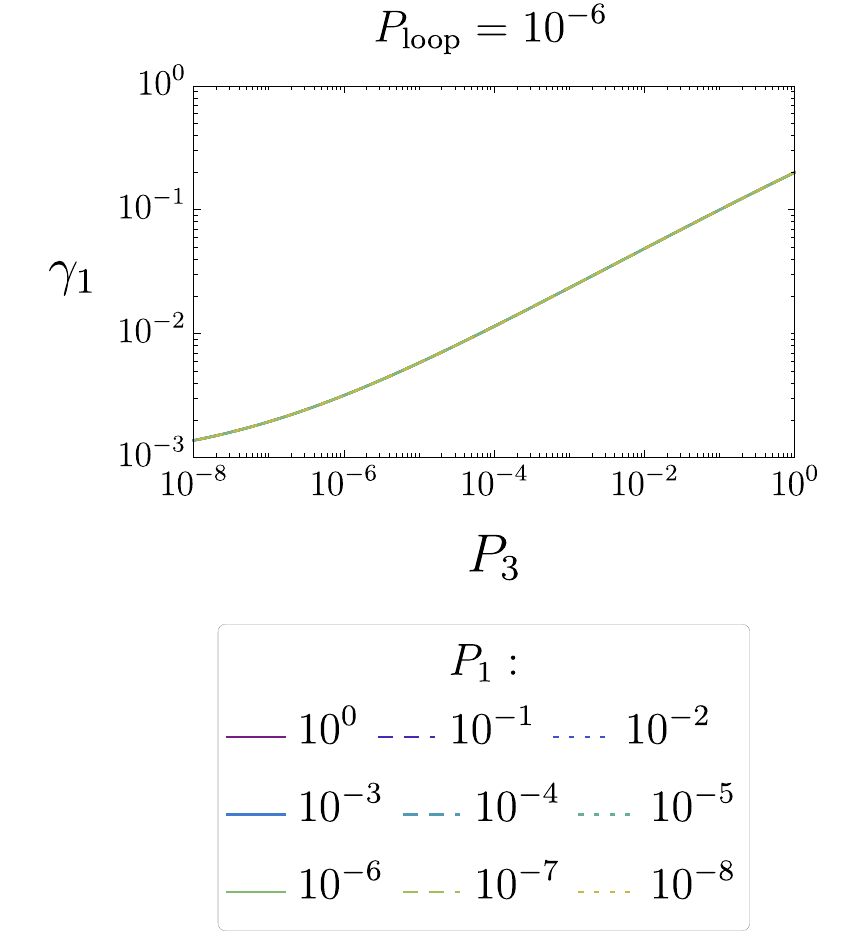}
  \end{minipage}
  \begin{minipage}{0.49\linewidth}
  \includegraphics[width=\linewidth]{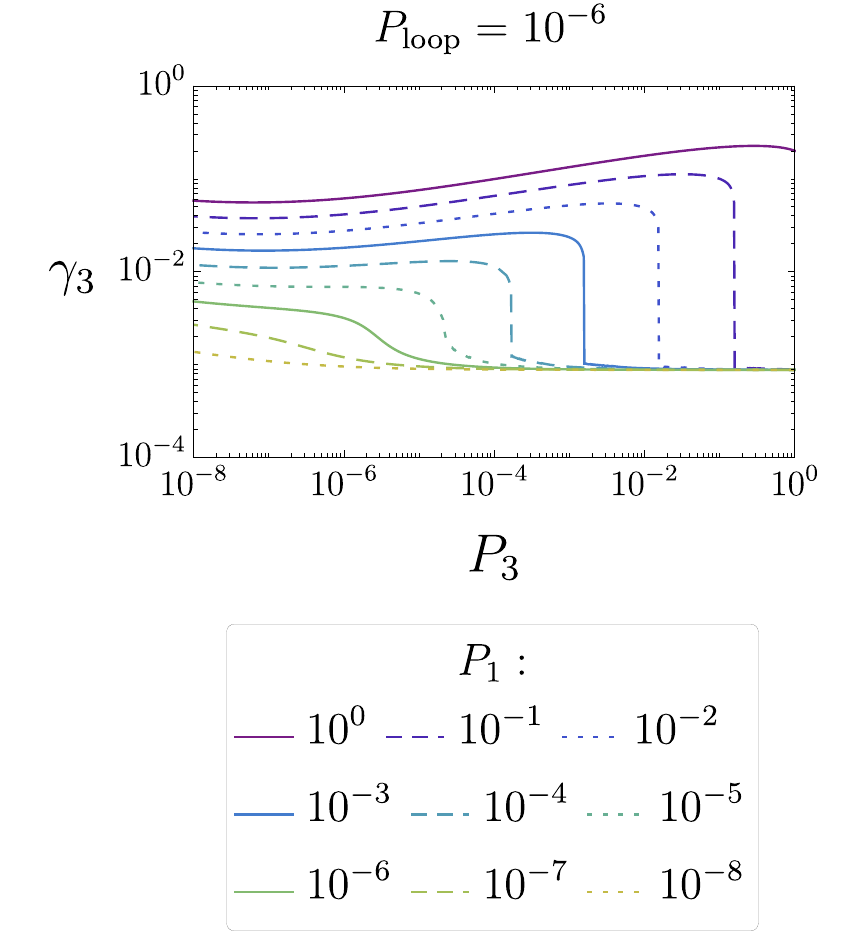}
  \end{minipage}
  \caption{Same as Fig.~\ref{plot-final value} but for the conventional VOS model. }
  \label{plot-final value forS}
\end{figure}

\begin{figure}
  \centering
  \begin{minipage}{0.48\linewidth}
  \includegraphics[width=\linewidth]{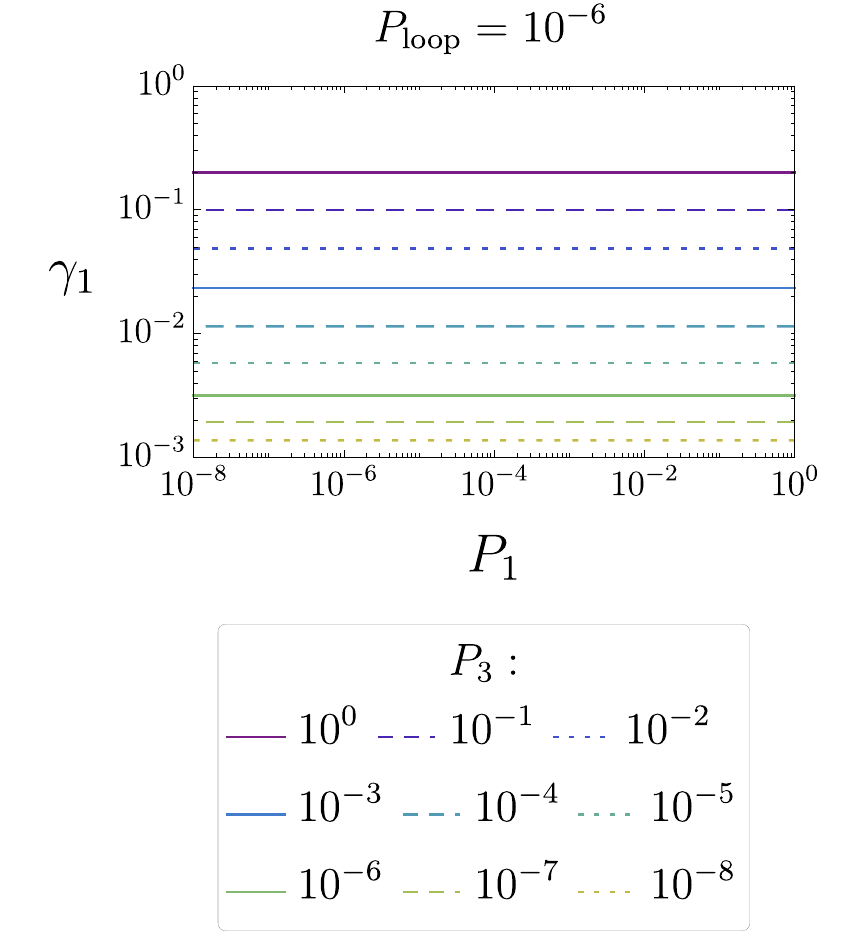}
  \end{minipage}
  \quad
  \begin{minipage}{0.48\linewidth}
  \includegraphics[width=\linewidth]{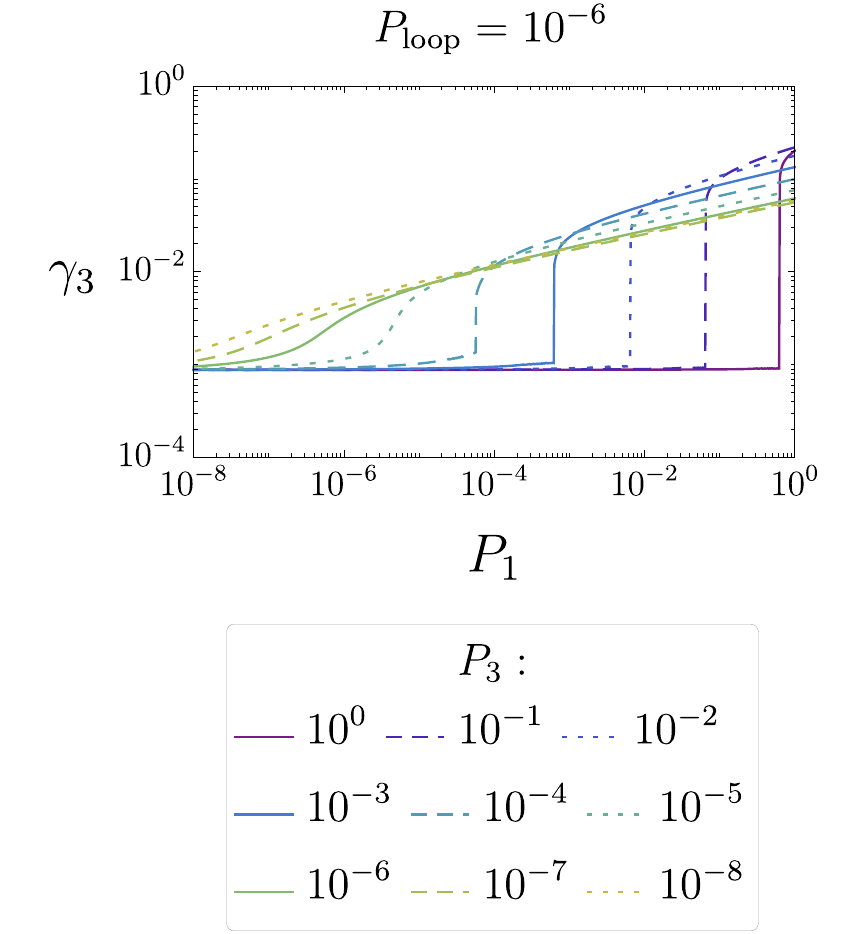}
  \end{minipage}
  \caption{Same as Fig.~\ref{plot-final value vs P1} but for the conventional VOS model. 
}
  \label{plot-final value vs P1 forS}
\end{figure}

\begin{figure}
  \centering
   \includegraphics[width=0.7\linewidth]{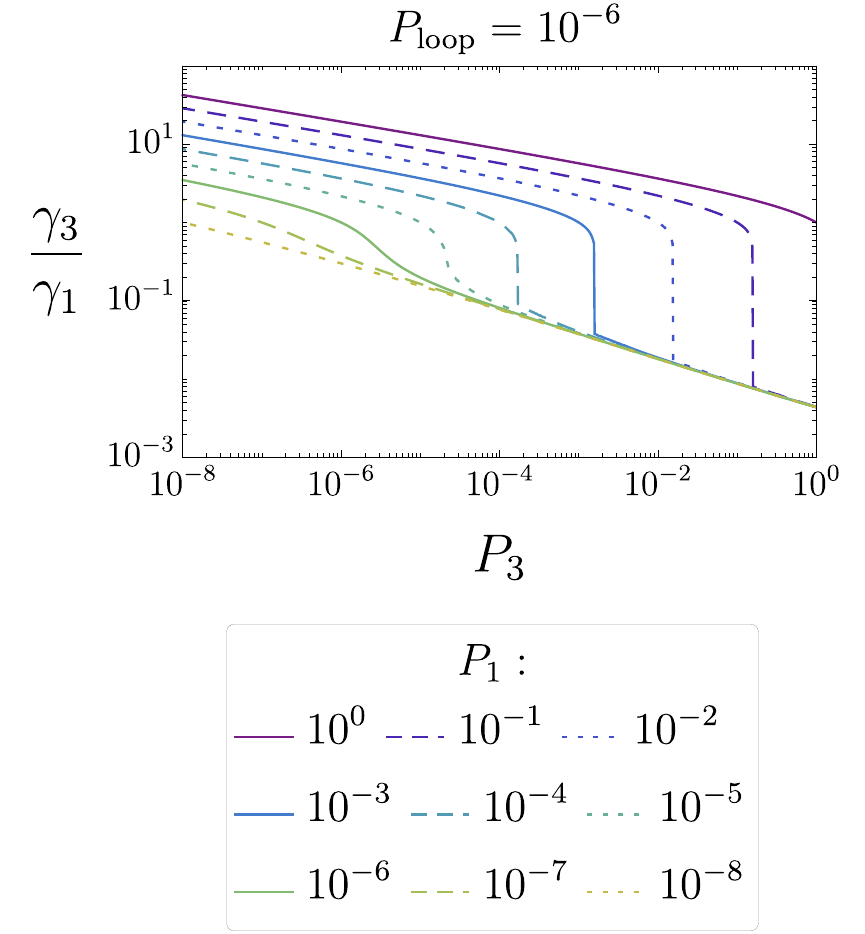}
  \caption{Same as Fig.~\ref{plot-ratio} but for the conventional VOS model. }
  \label{plot-ratio forS}
\end{figure}

Figures~\ref{plot-final value forS} and \ref{plot-final value vs P1 forS} show the scaling values of $\gamma_1$ (left panel) and $\gamma_3$ (right panel) as functions of $P_3$ and $P_1$, respectively, for $P_{\rm loop} = 10^{-6}$.
We vary $P_1 = P_2$ or $P_3$ from $10^{0}$ to $10^{-8}$, as indicated by different colors and line styles.
Figure~\ref{plot-ratio forS} shows the ratio $\gamma_3/\gamma_1$ as a function of $P_3$ for $P_{\rm loop} = 10^{-6}$.
Our numerical results for these scaling values are in good agreement with the theoretical predictions \eqref{eq:gamma1scalingcon1} and \eqref{eq:gamma1scalingcon2}.

\begin{figure}
  \centering
  \begin{minipage}{0.49\linewidth}
  \includegraphics[width=\linewidth]{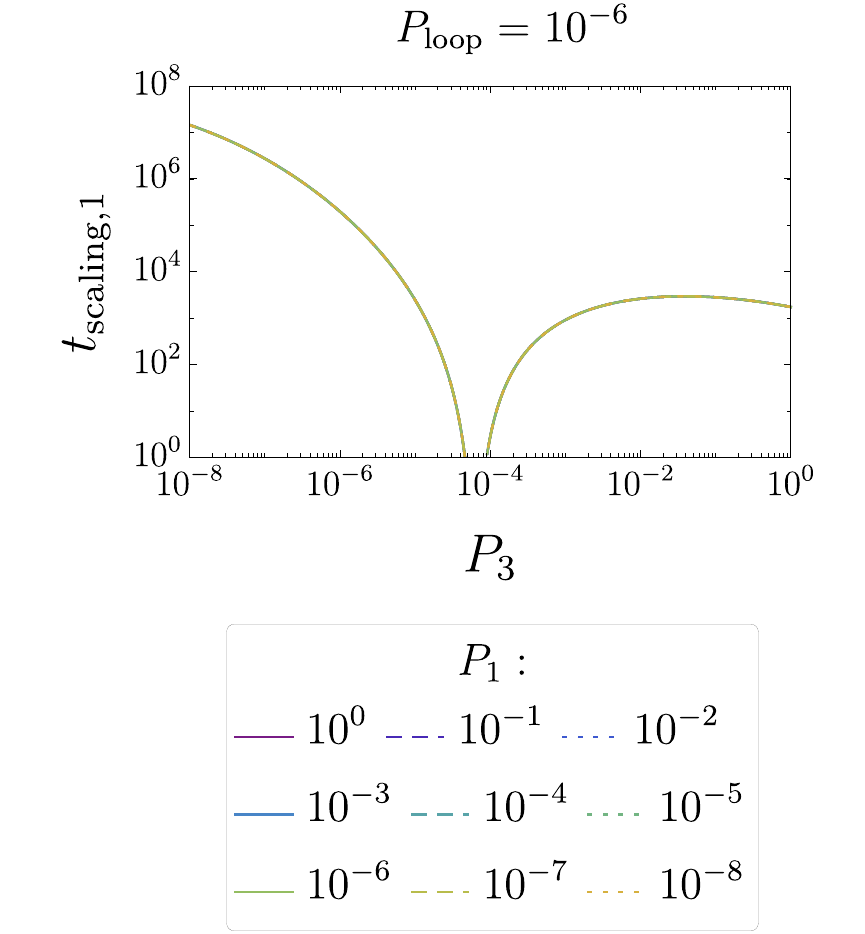}
  \end{minipage}
  \begin{minipage}{0.49\linewidth}
  \includegraphics[width=\linewidth]{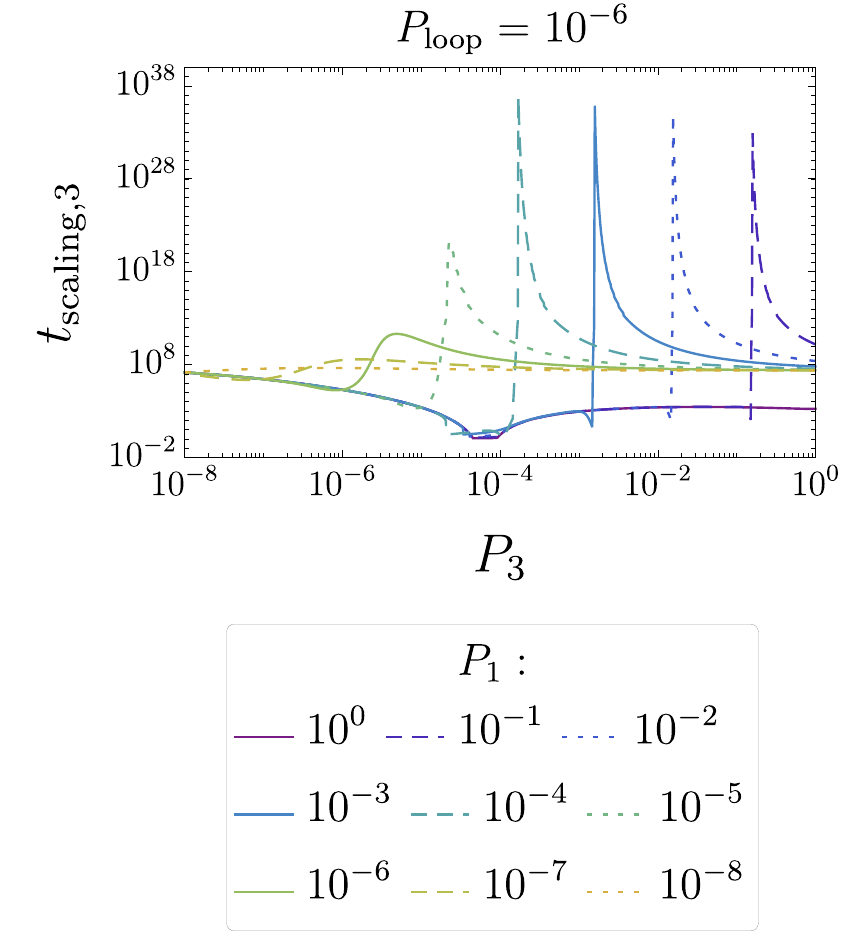}
  \end{minipage}
  \caption{Same as Fig.~\ref{plot-scaling time} but for the conventional VOS model. }
  \label{plot-scaling time forS}
\end{figure}

Figure~\ref{plot-scaling time forS} shows the scaling times $t_{\text{scaling},1}$ (left panel) and $t_{\text{scaling},3}$ (right panel) as functions of $P_3$ for $P_{\rm loop} = 10^{-6}$.
In the left panel, 
all results are nearly degenerate even when $P_1$ varies from $10^{0}$ to $10^{-8}$. 
In contrast,
we find that $t_{\text{scaling},3}$ exhibits a pronounced peak around specific values of $P_3$.
These values correspond to the parameter region in which the scaling solution for $\gamma_3$ changes abruptly, as seen in the right panels of Figs.~\ref{plot-final value forS} and \ref{plot-final value vs P1 forS}.
This behavior closely parallels that found in the extended VOS model.

\begin{figure}
\centering
   \begin{minipage}{0.48\linewidth}
    \centering
    \includegraphics[width=\linewidth]{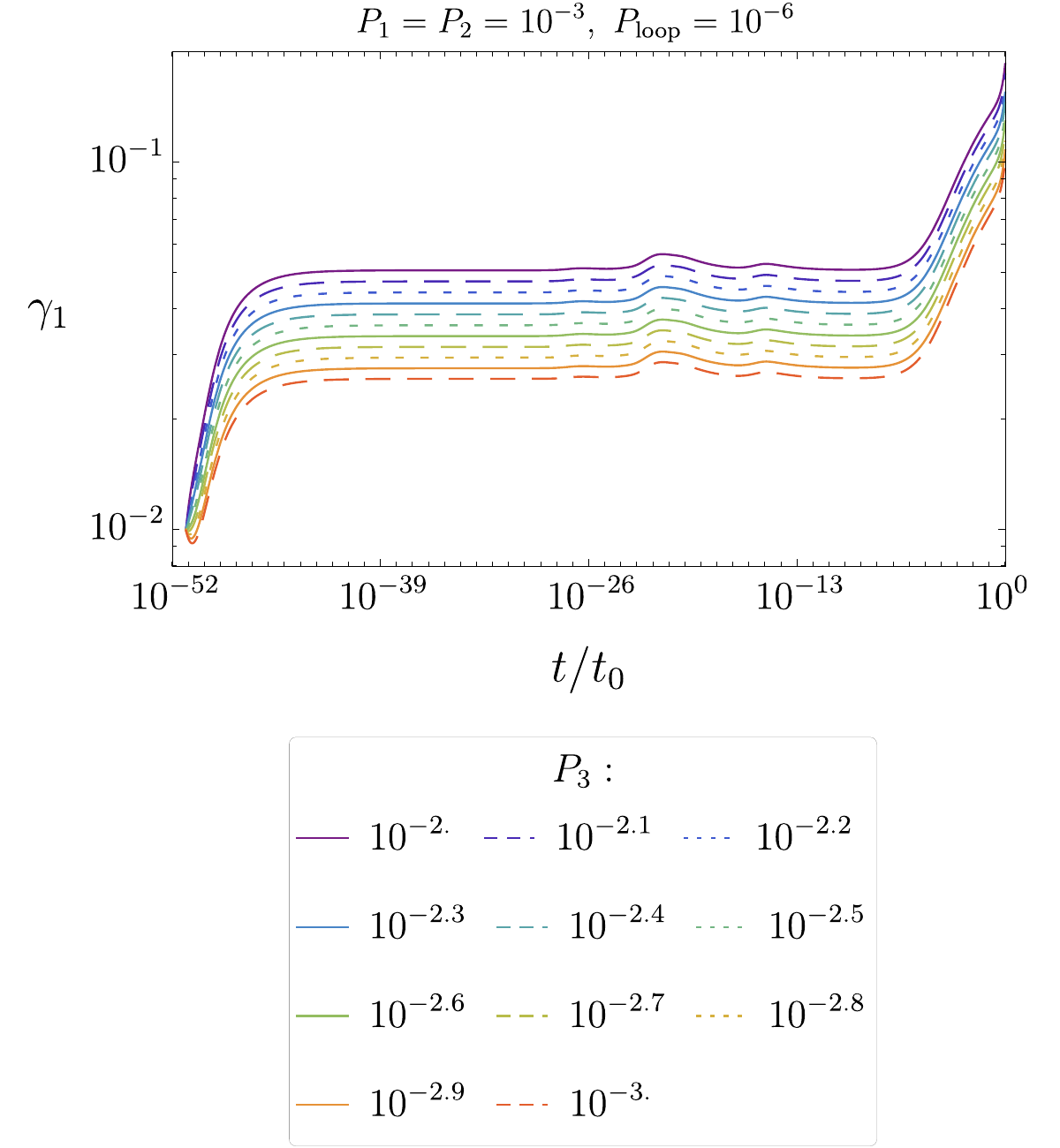}
  \end{minipage}
  \hspace{0.01\linewidth}
  \begin{minipage}{0.48\linewidth}
    \centering
    \includegraphics[width=\linewidth]{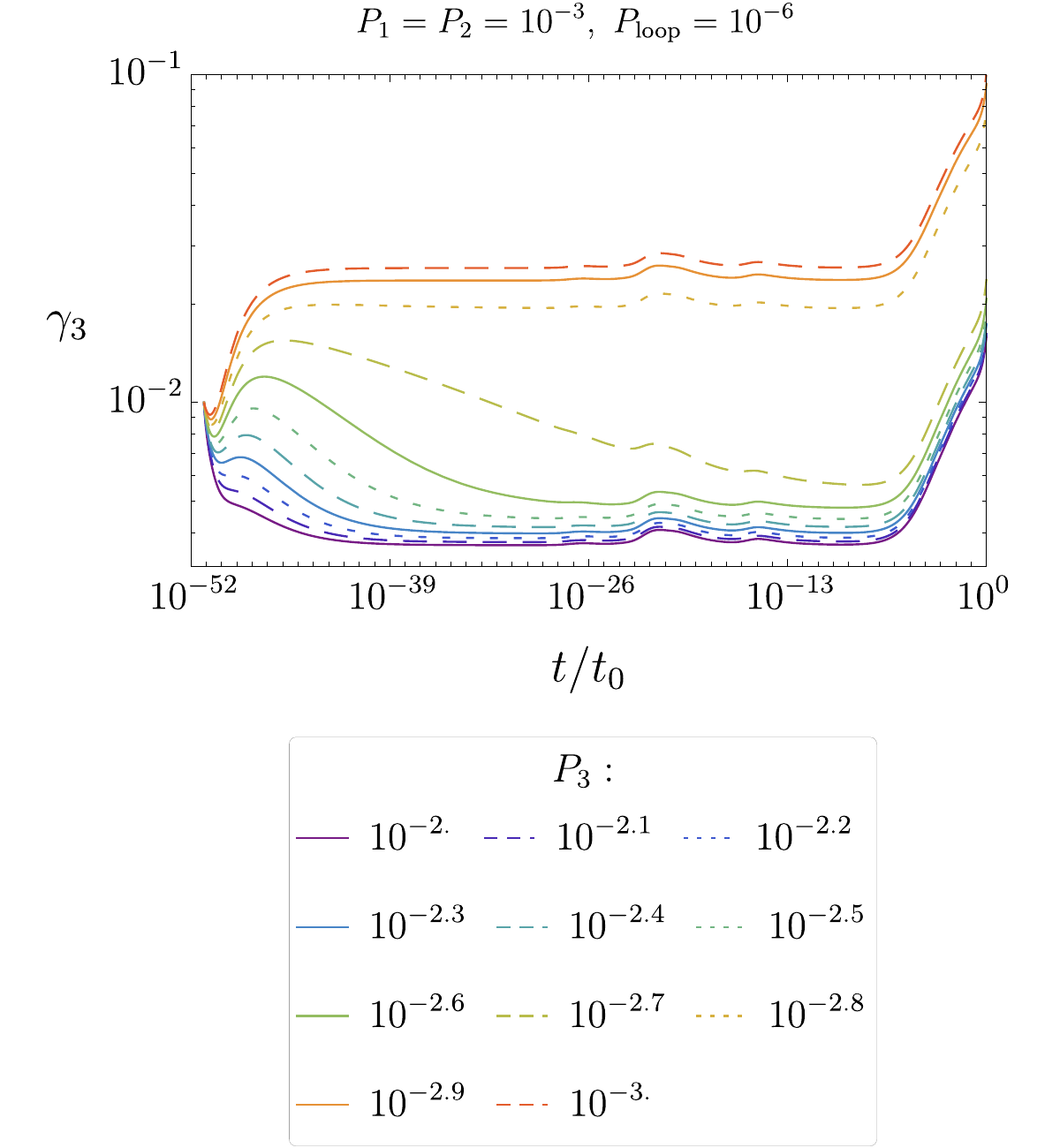}
  \end{minipage}
  \caption{Same as Fig.~\ref{timeevoluton for GW-delayscaling-byE} but for the conventional VOS model.}
  \label{timeevolution for GW-delayscaling-byS}
\end{figure}

\begin{figure}
    \centering
    \vspace*{0.2cm}
    \includegraphics[width=0.9\linewidth]{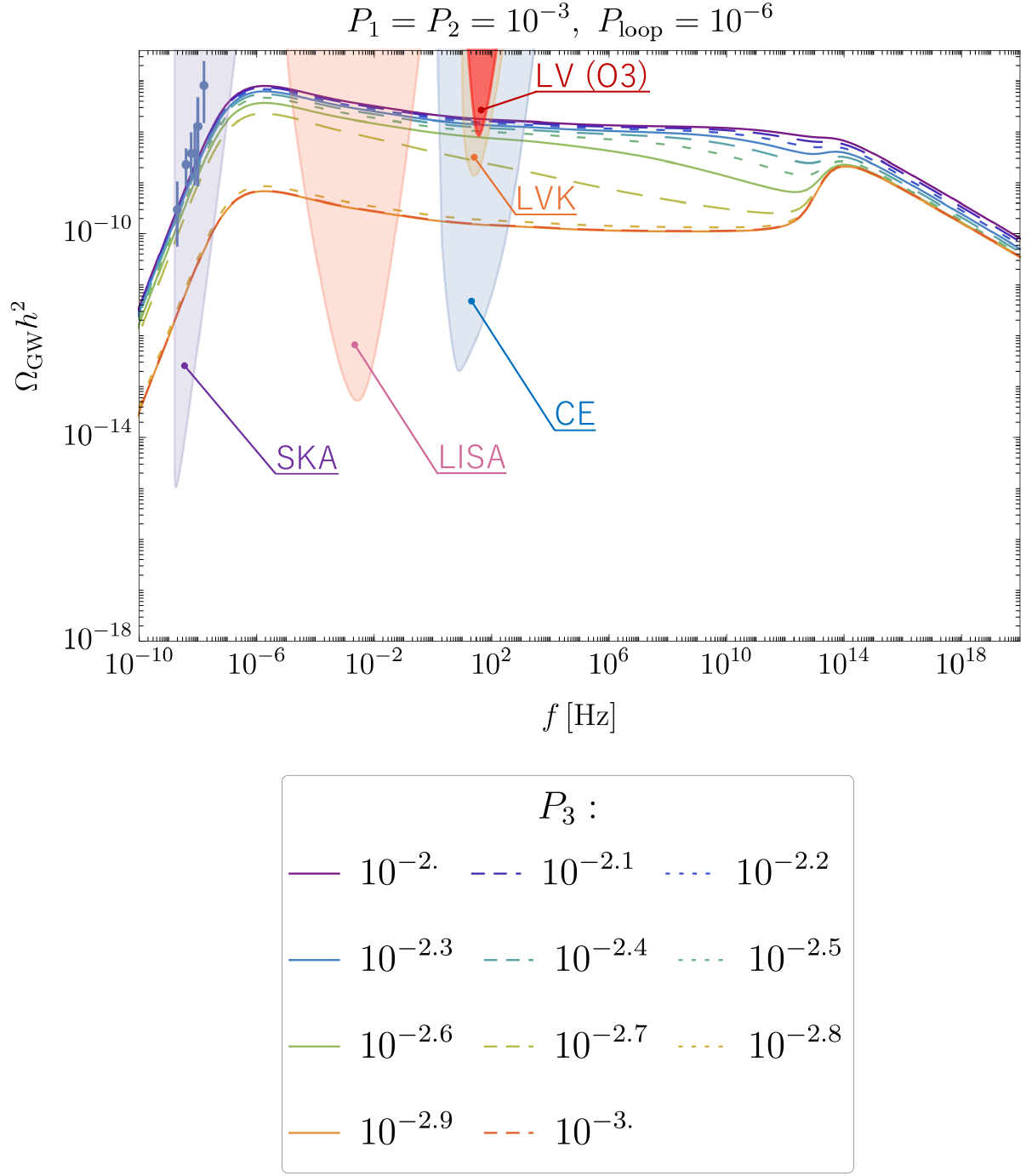}\\
    \caption{Same as Fig.~\ref{GW-delayscaling-byE} but for the conventional VOS model.}
  \label{GW-delayscaling-byS}
\end{figure}

\begin{figure}
  \centering
  \vspace*{0.2cm}
  \includegraphics[width=0.9\linewidth]{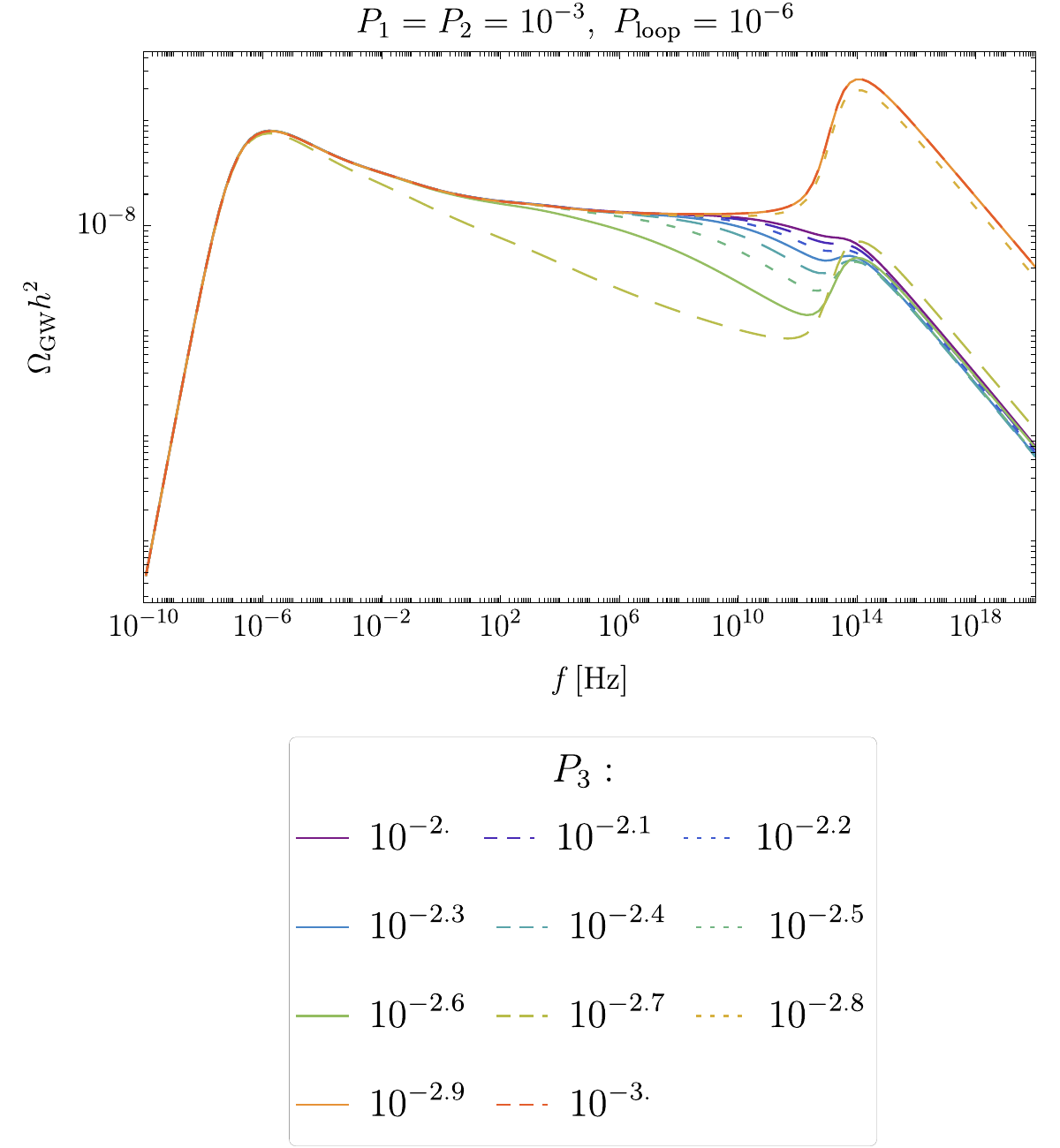}\\
  \vspace{0.7cm}
  \caption{Same as Fig.~\ref{GW-delayscaling-byS}, but with all spectra rescaled to coincide at low frequencies in order to highlight differences in the spectral shape.
  The overall normalization of the spectrum is not physically meaningful. 
}
  \label{freq dependence for delayscaling-byS}
\end{figure}

We next consider the realistic expansion history with \eq{eq:Hubble}.
Figure~\ref{timeevolution for GW-delayscaling-byS} shows the time evolution of $\gamma_1$ and $\gamma_3$ for a set of parameters chosen around those values that lead to a significantly delayed onset of the scaling regime.
Figure~\ref{GW-delayscaling-byS} presents the corresponding GW spectrum for the case $G\mu_i=10^{-12}$ for all $i$.
The impact of delayed scaling is more clearly illustrated in Fig.~\ref{freq dependence for delayscaling-byS}, where the GW amplitudes are rescaled to coincide in the low-frequency region.
As in the extended VOS case, a significantly delayed onset of the scaling regime enlarges the viable parameter space that can simultaneously account for the NANOGrav signal while remaining consistent with LIGO/Virgo constraints.

\begin{figure}
  \centering
  \vspace*{0.2cm}
  \includegraphics[width=0.9\linewidth]{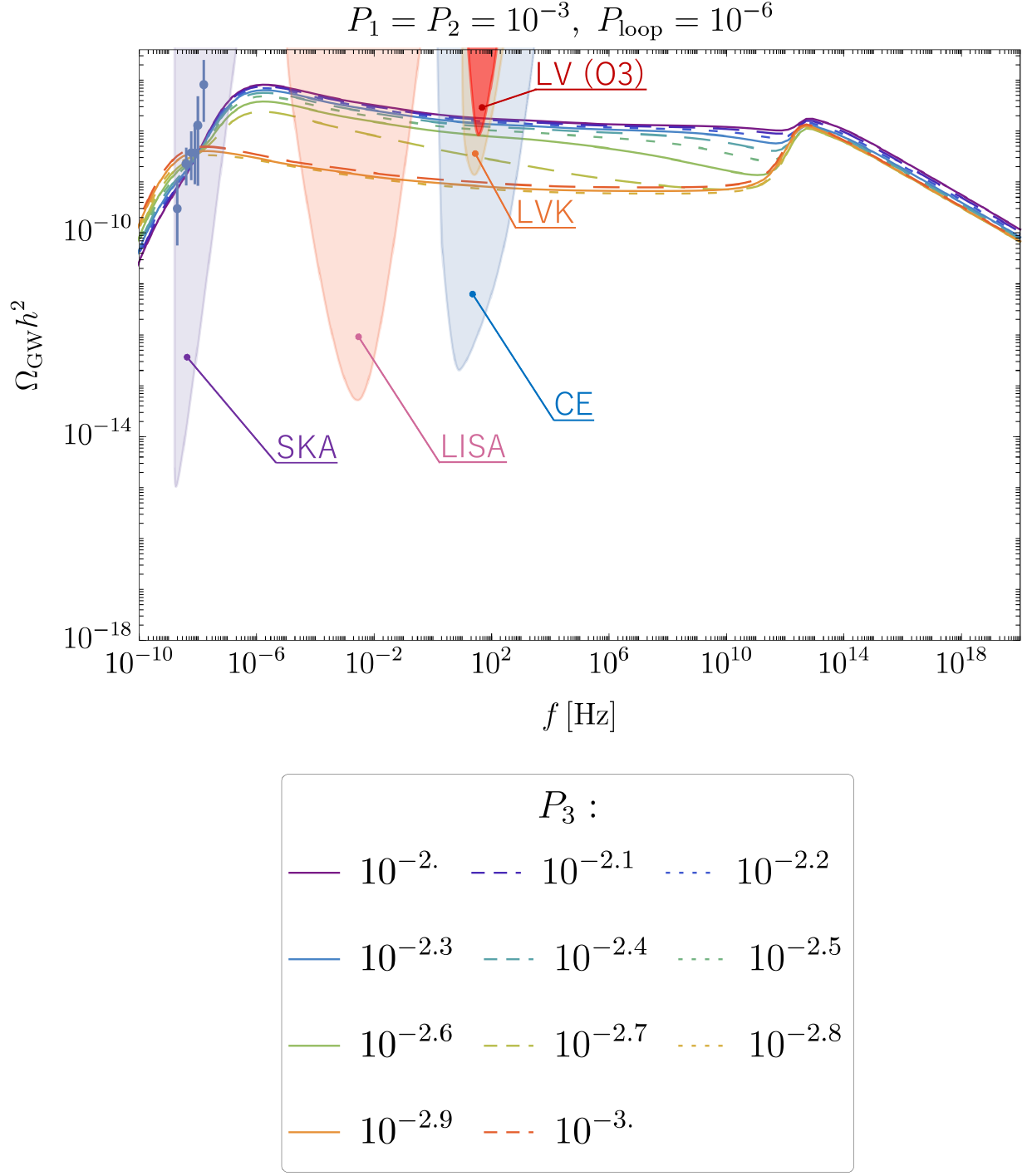}\\
  \vspace{0.7cm}
  \caption{
  Same as Fig.~\ref{GW-delayscaling-byE-Gmu} but for the conventional VOS model.}
  \label{GW-delayscaling-byS-Gmu}
\end{figure}

We further examine the GW spectrum for the case with a hierarchy of string tensions, while keeping all other parameters unchanged.
In particular, Fig.~\ref{GW-delayscaling-byS-Gmu} shows the GW spectrum for $G\mu_1=G\mu_2=10^2G\mu_3=10^{-10}$.
The resulting modifications to the GW spectrum in this hierarchical-tension scenario can be understood in the same manner as in the extended VOS model.

Overall, all of these qualitative features closely resemble those found in the extended VOS model.
In particular, for certain combinations of reconnection probabilities, the scaling time can become very long, and the resulting GW spectrum exhibits a suppression at high frequencies.

\section{Analytic calculation of scaling solution in conventional VOS model}
\label{sec:analyticforconventional}

Here we derive the analytic behavior of the scaling solution during RD in the conventional VOS model.

As discussed in Sec.~\ref{sec:B1}, the velocity should approach $v_i = 1/\sqrt{2}$ in order for $k(1/\sqrt{2}) = 0$ to hold when $\gamma_i \ll 1$.
Thus, when deriving the analytic scaling behavior, we may treat $v_i$ as a constant parameter.
For convenience, we also redefine
\begin{equation}
\begin{aligned}
    \tilde{P}_{\text{loop}}&:=cv(P_{\rm loop})^{\frac{1}{3}}\,,\\
    \tilde{P}_{\text{1}}&:=d\bar{v}c_{\xi}(P_1)^{\frac{1}{3}}\,,\\
    \tilde{P}_{\text{2}}&:=d\bar{v}c_{\xi}(P_2)^{\frac{1}{3}}\,,\\
    \tilde{P}_{\text{3}}&:=d\bar{v}c_{\xi}(P_3)^{\frac{1}{3}}\,,\\
\end{aligned}
\label{simplification of reconnection probability S}
\end{equation}
to simplify notation.

\subsection{Behavior of $\gamma_1$ }

In the scaling regime, we set $\dot\gamma_1 = 0$, which yields the following condition for the scaling solution:
\begin{align}
    0=(1+v^2)-2+\frac{\tilde{P}_{\text{loop}}}{\gamma_1}+\frac{\tilde{P}_3}{\gamma_2^2}\frac{\ell^3_{12}}{t}+\frac{\tilde{P}_2}{\gamma_3^2}\frac{\ell^2_{31}}{t}-\frac{\tilde{P}_1}{\gamma_3^2}\frac{\ell^1_{23}}{t} \,,
\end{align}
where we have substituted Eq.~\eqref{simplification3} into the evolution equation \eqref{seveq for gamma}.
Since $\gamma_1 = \gamma_2$, we have $\ell^2_{31} = \ell^1_{23}$, and therefore the last two terms cancel due to $P_1=P_2$ or $\tilde{P}_1=\tilde{P_2}$.
Furthermore, using $\ell^3_{12}/t = \gamma_1/2$, we obtain
\begin{align}
    0=-1&+\frac{\tilde{P}_\text{loop}}{\gamma_1}+\frac{\tilde{P}_3}{2\gamma_1}\label{scaling eq for gamma1}.
\end{align}
Solving this equation, the scaling value of $\gamma_1$ is
\begin{align}
    \gamma_1 = \tilde{P}_\text{loop}+\frac{1}{2}\tilde{P}_3.
\end{align}
Thus, the asymptotic behavior of $\gamma_1$ is
\begin{align}
\gamma_1 &=\mathcal{O}(1)\times
\begin{cases}
P_{\rm loop}^{\frac{1}{3}} & \text{for } P_3 \ll P_{\text{loop}} \vspace{0.3cm}
\\
P_3^{\frac{1}{3}} & \text{for }
P_{\text{loop}} \ll P_3 \,,
\end{cases}\label{gamma1 behavior}
\end{align}
where the dependence is expressed in terms of the original reconnection probabilities $P_{\rm loop}$ and $P_3$, not the redefined quantities $\tilde{P}_{\text{loop}}$ or $\tilde{P}_3$.

\subsection{Behavior of $\gamma_3$ }

In the same manner as for $\gamma_1$, we obtain the following equation that the scaling solution for $\gamma_3$ should satisfy:
\begin{align}
    0=(1+v^2)-2+\frac{\tilde{P}_{\text{loop}}}{\gamma_3}+\frac{2\tilde{P}_1\gamma_3}{\gamma_1(\gamma_1+\gamma_3)}-\frac{\tilde{P}_3\gamma_3^2}{2\gamma_1^3}\label{scaling eq for gamma3}.
\end{align}
This equation is more complicated than \eqref{scaling eq for gamma1} because it involves both $\gamma_1$ and $\gamma_3$.
We therefore analyze the behavior in different parameter regions.
In what follows, we approximate $(1+v^2)-2 \approx -1$.

\subsubsection{$\tilde{P}_3\ll \tilde{P}_\text{loop}$}

In this regime, we can approximate $\gamma_1 \approx \tilde{P}_\text{loop}$ from \eqref{gamma1 behavior}.
Substituting this into \eqref{scaling eq for gamma3}, we obtain
\begin{align}
    0=-1+\frac{\tilde{P}_{\text{loop}}}{\gamma_3}+\frac{2\tilde{P}_1\gamma_3}{\tilde{P}_{\text{loop}}(\tilde{P}_{\text{loop}}+\gamma_3)}
    -\frac{\tilde{P}_3\gamma_3^2}{2\tilde{P}_{\text{loop}}^3} \,.
\end{align}
Multiplying both sides by $\gamma_3(\tilde{P}_\text{loop}+\gamma_3)$ reduces the equation to the quartic form
\begin{align}
    \frac{\tilde{P}_3}{2\tilde{P}_\text{loop}^3}\gamma_3^4
    +\left(1-\frac{2\tilde{P}_1}{\tilde{P}_\text{loop}}\right)\gamma_3^2-\tilde{P}_\text{loop}^2 \approx 0.
\end{align}
where we use $\tilde{P}_3\ll \tilde{P}_\text{loop}$.

If $\tilde{P}_3\ll \tilde{P}_\text{loop}\ll \tilde{P}_1$, the equation can be approximated as 
\begin{align}
    &\frac{\tilde{P}_3}{2\tilde{P}_\text{loop}^3}\gamma_3^4 - \frac{2\tilde{P}_1}{\tilde{P}_{\rm loop}} \gamma_3^2-\tilde{P}_\text{loop}^2 \approx 0 \,. 
    \label{eq:P3loop1}
\end{align}
Solving this equation yields   
\begin{align}
    \gamma_3 \sim 
    2 \tilde{P}_1^{1/2}\tilde{P}_3^{-1/2}\tilde{P}_\text{loop}
    \propto \tilde{P}_1^{1/2}\tilde{P}_3^{-1/2}\tilde{P}_\text{loop} \,.
\end{align}
where the third term in \eq{eq:P3loop1} has been neglected as subdominant.

If $\tilde{P}_1, \tilde{P}_3 \ll \tilde{P}_\text{loop}$, we can approximate
\begin{align}
    \frac{\tilde{P}_3}{2\tilde{P}_\text{loop}^3}\gamma_3^4 + \gamma_3^2-\tilde{P}_\text{loop}^2 \approx 0\,. 
    \label{eq:P13loop}
\end{align}
Solving this equation yields
\begin{align}
    \gamma_3 \approx \tilde{P}_\text{loop} \,.
\end{align}
where the first term in \eq{eq:P13loop} has been neglected as subdominant.

\subsubsection{$\tilde{P}_\text{loop}\ll \tilde{P}_{3}$}

In this case, we can approximate $\gamma_1 \approx \tilde{P}_3$ from \eqref{gamma1 behavior}.
Then we obtain the following equation for the scaling value of $\gamma_3$ from \eqref{scaling eq for gamma3}:
\begin{align}
    \frac{1}{2\tilde{P}_3^2}\gamma_3^4+\frac{1}{2\tilde{P}_3}\gamma_3^3+\left(1-\frac{2\tilde{P}_1}{\tilde{P}_3}\right)\gamma_3^2+ \tilde{P}_3 \gamma_3-\tilde{P}_\text{loop}\tilde{P}_3\approx0.
\end{align}
For later convenience, it is useful to rewrite this equation as
\begin{align}
    \lkk \frac{1}{2} x^4 + \frac{1}{2} x^3+ x^2 + x \rkk -\frac{2\tilde{P}_1}{\tilde{P}_3} x^2  
     -\frac{\tilde{P}_\text{loop}}{\tilde{P}_3}=0.
    \label{gamma3 scaling equation for Ploop ll P3}
\end{align}
where we have defined $x\equiv \gamma_3/\tilde{P}_3$.

If $\tilde{P}_\text{loop}\ll \tilde{P}_3\ll \tilde{P}_1$, 
we first approximate the equation as 
\begin{align}
    \frac{1}{2} x^4 -\frac{2\tilde{P}_1}{\tilde{P}_3} x^2 \approx 0.
    \label{eq:Ploop31}
\end{align}
Solving this equation yields
\begin{align}
    \gamma_3 \approx 2 \tilde{P}_1^{1/2}\tilde{P}_3^{1/2} \propto&~ \tilde{P}_1^{1/2}\tilde{P}_3^{1/2} \,, 
\end{align}
which self-consistently justifies the approximation in \eqref{eq:Ploop31} $\tilde{P}_\text{loop}\ll \tilde{P}_3\ll \tilde{P}_1$.

If $\tilde{P}_\text{loop}\ll \tilde{P}_1 \ll \tilde{P}_3$ or $\tilde{P}_1\ll \tilde{P}_\text{loop} \ll \tilde{P}_3$, the equation reduces to 
\begin{align}
    \lkk \frac{1}{2} x^4 + \frac{1}{2} x^3+ x^2 + x \rkk  
     -\frac{\tilde{P}_\text{loop}}{\tilde{P}_3}=0.
\end{align}
Assuming $x \ll 1$, this equation admits the approximate solution
\begin{align}
 x \approx \frac{\tilde{P}_\text{loop}}{\tilde{P}_3}, 
\end{align}
which self-consistently justifies the assumption $x \ll 1$.
We therefore obtain
\begin{align}
    \gamma_3 
    &\propto \tilde{P}_\text{loop}\,,
\end{align}
in this regime.

In summary, the asymptotic behavior of $\gamma_3$ is 
\begin{align}
    \gamma_3 &=\mathcal{O}(1)\times
\begin{cases}
P_{\rm loop}^{1/3} & \text{for }\quad
P_1 \ll {\rm Max}[ P_{\rm loop}, \,  P_3  ]
\vspace{0.3cm}\\
P_1^{1/6} P_3^{1/6}& \text{for}\quad
P_{\rm loop} \ll P_3 \ll  P_1 
\vspace{0.3cm}\\
P_1^{1/6}P_3^{-1/6}P_{\rm loop}^{1/3} & \text{for}\quad
P_3 \ll P_{\rm loop}\ll P_1\,, \\
\end{cases}
\end{align}
where the dependence is expressed in terms of the original reconnection probabilities $P_{\rm loop}$, $P_1$, and $P_3$ not the redefined quantities $\tilde{P}_{\text{loop}}$ or $\tilde{P}_i$.

\section*{Acknowledgements}
This work was supported by JSPS KAKENHI Grant Numbers 23K13092 [MY].

\bibliography{reference}

\end{document}